\documentclass[journal,12pt,draftclsnofoot,onecolumn]{IEEEtran}

\usepackage{amsfonts}
\usepackage{amsmath}

\usepackage{cite}      

\usepackage{graphicx}  

\usepackage{stfloats}  

\newcommand{\comment}[1]{}
\newcommand{\mycomment}[1]{}
\newcommand{\obmcmntout}[1]{} %
\newcommand{\obmreuse}[1]{} %
\newcommand{\obmsubmittodo}[1]{} %

\newcommand{\obmabrdgol}[1]{} %
\newcommand{\obmthg}[1]{} 

\newcommand{\figcomment}[1]{}

\begin{document}

\title{Going Towards Discretized Spectrum Space: Quantification of Spectrum Consumption Spaces and a Quantified Spectrum Access Paradigm}

\author{Nilesh~Khambekar,~\IEEEmembership{Member,~IEEE,}
        Chad~M.~Spooner,~\IEEEmembership{Senior Member,~IEEE,}
        and~Vipin~Chaudhary,~\IEEEmembership{Member,~IEEE}
\thanks{Nilesh Khambekar and Vipin Chaudhary are with the Department of Computer Science and Engineering, University at Buffalo, SUNY, Buffalo, New York 14260-2000}
\thanks{Chad M. Spooner is with NorthWest Research Associates.}}


%


\maketitle

\begin{abstract}
Dynamic spectrum sharing approach is a paradigm shift from the conventional static and exclusive approach to spectrum allocation. The existing methodologies to define use of the spectrum and quantify its efficiency are based on the static spectrum assignment paradigm and not suitable for the dynamic spectrum sharing paradigm. There is a need to separately quantify the spectrum consumed by the individual transmitters and receivers when multiple heterogeneous wireless networks are sharing the spectrum in time, space, and frequency dimensions. By discretizing the spectrum dimensions, we define a methodology for quantifying the spectrum consumption spaces. This is an attempt to adopt the discretized signal processing principle and apply it to spectrum management functions that would bring in simplicity, flexibility, and precision among other advantages.
\end{abstract}


%
\IEEEpeerreviewmaketitle



\section{Introduction}


Spectrum sharing among multiple heterogeneous wireless networks is becoming increasingly important as the demand for the limited radio frequency spectrum is growing. Static allocation of spectrum results in severe underutilization in the time, space, and frequency dimensions \cite{fcc_sptf}. A dynamic spectrum access enables sharing of spectrum among multiple networks in the time, space, and frequency dimensions. 

To be able share spectrum, we need to identify the what spectrum is used and what spectrum can be shared with cochannel networks (usually without causing harmful interference). Here, we contrast the dynamic spectrum sharing paradigm with the static and exclusive spectrum-usage paradigm. In case of static and exclusive spectrum-usage paradigm, the usage of spectrum is generally captured in terms of RF-power from the network transmitter(s). As we transit to dynamic spectrum sharing paradigm, we need to the ability to identify, quantify, estimate the spectrum that can be shared. 

We note that there exist multiple spectrum sharing mechanisms that primarily differ in terms of interference management approach. For example, Opportunistic Spectrum Access (OSA) mechanism permits secondary access to spectrum when the primary network is not operating while Underlay spectrum access mechanism allows accessing spectrum even when the primary network is operating but the secondary transmit-power is very much constrained. In this regard, we need the ability to decouple the spectrum access mechanism that \textit{implies} certain spectrum availability based on its aggressive or conservative interference management approach and  instead quantify the \textit{absolute} utilized spectrum and the \textit{absolute} available spectrum.

Traditionally, it is assumed that spectrum is consumed by transmitters only; However, receivers \textit{also} consume by receivers by denying or constraining the transmit-power by other transmitters \cite{itu}. In case of exclusive spectrum-usage paradigm, the  spectrum consumed by receivers need not be separately considered. For quantifying spectrum consumption under spectrum sharing paradigm, we argue that the spectrum consumed by the individual transmitters and receivers needs to be considered. \mycomment{While the RF-power received from a transmitter decreases with increasing distance from the transmitter; the constraint imposed by a receiver on the tolerable interference power increases with increasing distance from the receiver. In order to \textit{quantify} the spectrum consumed by transmitters and receivers,} In this regard, we propose to discretize the time, space, and frequency dimensions and incorporate the absolute maximum and absolute minimum power limits at any point in space. The concept of spectrum-space discretization enables us to characterize and quantify multiple spectrum spaces. The quantification of spectrum consumption spaces quantifying, analyzing, and comparing performance of spectrum recovery and exploitation mechanisms. Furthermore, it facilitates a quantified approach to spectrum management and spectrum access regulation.

The paper is organized as follows. In Section 2, we describe the limitations of the traditional approaches to quantifying spectrum consumption. In Section 3, we present the methodology for quantifying spectrum consumption spaces. In this regard, we describe discretization of spectrum space. We also discuss considerations while discretizing the spectrum space dimensions. In Section 4, we illustrate the quantification methodology with example use cases. In Section 5, we discuss spectrum management functions in the context of spectrum space discretization. In Section 6, we discuss how spectrum space discretization enables analysis and optimization of the spectrum management functions. Based on quantification of spectrum consumption spaces, we describe a \textit{quantified dynamic spectrum access paradigm} that facilitates real time dynamic spectrum sharing in Section 7. Finally, we draw conclusions and outline future research avenues in Section 8. 

\section{Motivation and Related Work}

\subsection{Traditional Metrics for Quantifying Spectrum Usage}
The metrics being used today for performance analysis of wireless networks are not well suited for analyzing the spectrum consumption performance of spectrum access mechanisms. International Telecommunication Union (ITU) has defined the \textit{spectrum utilization factor} as the product of the bandwidth, geometric space, and the time denied to other potential users \cite{itumetrics}. However, spectrum utilization factor does not represent \textit{actual} usage at the system level. 

ITU defined \textit{spectrum utilization efficiency} (SUE) as a ratio of the amount of information transferred to the spectrum utilization factor \cite{itumetrics}. This performance metric is defined to measure the performance in the context of \textit{a wireless system} and cannot be applied for analyzing spectrum consumption performance of \textit{a spectrum access mechanism} that allows multiple wireless networks to coexist together. 

Federal Communications Commission (FCC) introduced a metric \textit{interference temperature} to enforce the existing spectrum access rights of incumbents and enable dynamic spectrum access \cite{fccmetrics}. \textit{The interference temperature is a measure of the RF power available at a receiving antenna to be delivered to a receiver - power generated by other emitters and noise sources.} The interference temperature of a receiver specifies \textit{the worst case} environment in which it can operate. While the interference temperature metric was suggested for the protection of spectrum-rights of incumbents, it cannot be directly used for quantifying spectrum consumption.

`XG Technology' defined \textit{white space fill factor} and \textit{success in channel use} for measuring spectrum efficiency \cite{xgmetrics}. The \textit{success in channel use} metric depends upon the number of networks contending for a set of channels. The \textit{white space fill factor} metric is based on MAC efficiency of a network and thus cannot be applied for the whole system of wireless networks. XG used \textit{abandon-time} and \textit{interference-to-noise ratio} metrics for ensuring no harm by XG networks to the incumbents. This metric cannot quantity the harmfully interfered spectrum in the spatial and temporal dimensions.

With regards to evaluating the performance of \textit{recovering} the underutilized spectrum, the sensitivity and ROC plots of the detector are suited for comparing performance of detection algorithms. Tandra et. al. defined safety and performance metrics for comparing detection algorithms \cite{tandra_metrics}. However, these metrics cannot be used for \textit{quantifying} the spatial, spectral and temporal spectrum recovered by employing a spectrum sensing technique. 

\subsection{Related Work}
Based on the concept of usage of spectrum by receivers \cite{itumetrics}, we observed the need to separately quantify the spectrum consumed by individual transmitters and receivers when multiple heterogeneous wireless networks are sharing the spectrum in time, space, and frequency dimensions. By discretizing the spectrum dimensions, we defined a methodology to quantify spectrum consumption spaces. 

Stine has proposed an approach to model the spectrum-access parameters of transmitters, receivers, and RF systems and suggested the use of models for spectrum management \cite{crt_stine}. This approach is primarily focused on defining standard models and assessing the compatibility between uses of spectrum. The emphasis of the proposed approach is on absolute quantification of spectrum consumption spaces that can be applied in analysis, design, and comparison of spectrum management functions.

In order to improve spectrum sharing, it is important to understand the weaknesses and quantify their impact. The poor exploitation of the spectrum due to conservative spectrum-access constraints is illustrated in \cite{berk_wsc}. In \cite{oms2_sca}, we have quantified the loss of the available spectrum due to the lack of knowledge of the RF-environment. 

To maximize the recovery of the available spectrum, it is necessary to acquire the RF environment information.  In \cite{sgn_icnc}, we described algorithms to estimate the RF environment information exploiting signal cyclostationarity. Similar to `Sensing as a Service' \cite{saas_weiss}, we separate the sensing function from the secondary user radio and apply an external sensor network based infrastructure for estimating spectrum consumption in real time\cite{oms4_sce}. 

We addressed the problem of efficient exploitation of the available spectrum in \cite{oms3_cf1}. In this regard, we introduced the quantified spectrum access paradigm, suggested an efficient approach to scheduling spectrum-access requests and evaluated various design choices for a spectrum access mechanism (SAM). 

\section{Spectrum Consumption Quantification Methodology}

\subsection{Approach}
We consider a generic system of multiple heterogeneous wireless networks. In order to capture the lowest granularity of spectrum-use in the system, we consider a network represents a spectrum-access request with a transmitter and one or more receivers. 

\noindent
\textbf{Interference Model}
The power received from a transmitter $t_n$ at a point $\rho$ in the spatial dimension is given by 
\begin{equation}
\label{eq:gen_rcvd_power}
P_{r_{\rho}}(t_n) = P_{t_n} min\Big\{1, L({d(t_n,\rho)}^{\alpha})\Big\}
\end{equation}
where $P_{t_n}$ is the transmit power of the transmitter and $d(t_n,\rho)$ is the distance between the transmitter $t_n$ and the point $\rho$ in the space. ${\alpha}$ is the path-loss exponent and it is assumed that $\alpha > 2$.  $L({d(t_n,\rho)}^{\alpha})$ denotes the path-loss factor. The \textit{min} operation in ~(\ref{eq:gen_rcvd_power}) ensures that the received power is never more than the transmitted power.

We assume that the transceivers employ directional transmission and reception in order to minimize interference. A receiver can withstand certain interference when the received Signal to Interference and Noise Ratio (SINR) is greater than the given threshold, $\beta$\footnote{The threshold, $\beta$, represents the quality of a receiver and incorporates receiver-noise and other receiver technology imperfections.}. 

We define the maximum power at any point in the system to be $P_{MAX}$ and the minimum power be $P_{MIN}$. In practice, the transmitter design factors and the human safety conditions determine $P_{MAX}$ and $P_{MIN}$ is driven by minimum possible measurable power. $P_{MAX}$ and $P_{MIN}$ together enable us to quantify the spectrum consumed by a transmitter or receiver in \textit{absolute} terms.

\noindent
\textbf{Discretizing Spectrum Access Dimensions}

In order to quantify the spectrum consumption by a transmitter over a range of time, space, and frequency, we need to \textit{sample} the signal in the time, space, and frequency dimensions and calculate the average utilized spectrum.

In practice, we consider transmissions over a band of frequencies and therefore we compute the spectrum consumption in terms of unit bandwidth. The spectrum consumed by the transmitter in a unit bandwidth is the \textit{average power} of the signal over the range of frequencies in the unit bandwidth n the given range of time and space dimensions.

Similarly, we discretize the amount of time a transmitter is exercising the access to the spectrum and express in terms of the number of unit time-quanta.
The spectrum consumed by the transmitter in a unit time quantum is the \textit{average power} of the signal over the unit time quantum, in the given range of frequency and space dimensions. 

Finally, we discretize the geographical region under interest into unit-regions. The spectrum consumed by the transmitter in a unit region is the \textit{average power }of the signal \textit{over the unit region} in the given range of time and frequency dimensions.

In the next subsection, we quantify spectrum consumption by transmitters and receivers considering the discrete spectrum access dimensions. 

\subsection{Quantifying Spectrum Consumption in a Discrete Spectrum Space}

In this subsection, we first quantify spectrum is consumed by transmitters and receivers at a point in a discrete grid of points in spectrum space. We identify different spectrum consumption spaces depending on how spectrum is consumed at a point. Finally, we combine the spectrum consumption at the discrete points and quantify those spectrum consumption spaces.

\subsubsection{Spectrum consumption at a point}
The RF-power received from a transmitter at a point is given by eq.(\ref{eq:gen_rcvd_power}).

\noindent
\textbf{Spectrum occupancy at a point} \\
The aggregate power received at a point $\rho$ in space, in a spectral band, at a given time is defined as the spectrum-occupancy, $\omega(\rho)$. It is given by
\begin{equation}
\label{eq:spocpt}
\omega(\rho) = \sum_{n}	P_{r_\rho}(t_n) + W_{\rho} ,
\end{equation}
where $W_{\rho}$ is the average ambient noise power at $\rho$. The unit for spectrum-occupancy is $W$.

\noindent
\textbf{Spectrum opportunity at a point} \\
We define the maximum additional power that could be emitted from a point without causing harmful interference to \textit{any} of the cochannel receivers as the \textit{spectrum-opportunity} at that point.

Let $r_{n,m}$ be the $m^{th}$ receiver of the $n^{th}$ network. The amount of interference receiver $r_{n,m}$ can tolerate, termed interference-margin, is 
\begin{equation}
IM_{r_{n,m}, t_n} = \frac{P_{r_{n,m}}(t_n)}{\beta_{n,m}} - W_{r_{n,m}} .
\end{equation}
The unit of interference-margin is $W$.

We can view interference-margin $IM_{r_{n,m}, t_n}$ as the upper-bound on the transmit-power of an interferer at a spatial separation of zero. We quantify the limit on interference-power at a point $\rho$ in space in terms of the receiver-imposed interference-power upper bound.
\begin{equation}
\label{eq:riub}
\acute{I}(r_{n,m}, t_n, \rho) = IM_{r_{n,m}} min\{1, L(d(\rho, r_{n,m})^{\alpha})\} ,
\end{equation}
where $d(\rho, r_{n,m})$ is the distance between the receiver $r_{n,m}$ and the point $\rho$ in the space. 

Let $\breve{I}(r_{n,m}, t_n, \rho)$ represent the aggregate interference seen by receiver $r_{n,m}$. Then, the \textit{interference opportunity} imposed by this receiver is given by the difference between the upper bound on the interference and the aggregate interference experienced.
\begin{equation}
\label{eq:spoppt}
\ddot{I}(r_{n,m}, t_n, \rho) = \acute{I}(r_{n,m}, t_n, \rho) -  \breve{I}(r_{n,m}, t_n, \rho) .
\end{equation}

By spatially combining the limits on the maximum interference power imposed by all the receivers, from all the networks in the system, we obtain \textit{spectrum-opportunity} at a point $\rho$ in the unit-region as
\begin{equation}
\gamma(\rho)	= \min_n (\min_m (\ddot{I}(r_{n,m}, t_n, \rho))) .
\end{equation}
The unit of spectrum-opportunity at a point is $W$.

\noindent
\textbf{Receiver-liability at a point} \\
The receivers impose a limit on the maximum power at a point for successful reception. We define receiver-liability as the maximum power \textbf{\textit{denied}} to the potential transmitters by the cochannel receivers at a point in a spectral band at a given time.

The maximum power allowed by the cochannel receivers at a point is the sum of transmitter occupancy and interference opportunity. The receiver-liability is the difference between the maximum power at a point and the maximum power allowed by the cochannel receivers.
\begin{equation}
\label{eq:sprlpt}
\phi(\rho) = P_{MAX} - (\omega(\rho) + \gamma(\rho)) ,
\end{equation}
The unit of spectrum-opportunity at a point is $W$.

\subsubsection{Spectrum Consumption in the Discrete Spectrum Space}
Next, we quantify 
\begin{itemize}
	\item utilized spectrum: the spectrum consumed by the \textit{transmitters} in the time, space, and frequency dimensions.
	\item forbidden spectrum: the spectrum consumed by the \textit{receivers} in the time, space, and frequency dimensions..
	\item available spectrum: the spectrum that \textit{can be availed} by future spectrum-accesses in the time, space, and frequency dimensions.
\end{itemize}

\noindent
\textbf{The Total Spectrum Space}

Let the geographical region be discretized into $\hat{A}$ unit-regions, $\hat{B}$ unit-frequency-bands, and $\hat{T}$ unit-time-quanta.  Thus, the total spectrum space is given by
\begin{equation}
\label{eq:p1totalspectrum}
\Psi_{Total} = P_{MAX} \hat{T} \hat{A} \hat{B} .
\end{equation}
The unit of the total spectrum is $Wm^2$. 

\noindent
\textbf{Utilized Spectrum in a geographical region ($\Psi_{utilized}$)}
We define \textit{spectrum-occupancy} in a unit-region $\chi$, in the ${\tau}^{th}$ time-quantum, and in the frequency-band $\nu$ as the aggregate power received at a sample point $\rho_0 \in \chi$ in the unit-region. 
\begin{equation}
\label{eq:urspoc}
\Omega(\chi, \tau, \nu) = \omega({\rho_0}, \tau, \nu) .
\end{equation}
The spectrum utilized in a geographical region is the sum of the spectrum-occupancy in all the unit-regions across the time and frequency dimensions.
\begin{equation}
\label{eq:agsput}
\Psi_{utilized} = \sum_{k=1}^{\hat{B}} \sum_{j=1}^{\hat{T}} \sum_{i=1}^{\hat{A}} {\Omega}(\chi_i, \tau_j, \nu_k) ,
\end{equation}

\noindent
\textbf{Available Spectrum in a geographical region ($\Psi_{available}$)}

The \textit{spectrum-opportunity} in a unit-region $\chi$ at the ${\tau}^{th}$ snapshot of time, and in the frequency-band $\nu$ is defined as the spectrum-opportunity at a sample point $\rho_0 \in \chi$ in the given unit-region. 
\begin{equation}
\label{eq:urspop}
{\Gamma}(\chi, \tau, \nu) = \gamma(\rho_0, \tau, \nu) .
\end{equation}
To quantify the available-spectrum in a geographical region, we need to sum spectrum-opportunity in all the unit-regions across the temporal and spectral dimensions
\begin{equation}
\label{eq:agspav}
\Psi_{available} = \sum_{k=1}^{\hat{B}} \sum_{j=1}^{\hat{T}} \sum_{i=1}^{\hat{A}} {\Gamma}(\chi_i, \tau_j, \nu_k) .
\end{equation}

\noindent
\textbf{Forbidden spectrum in a geographical region ($\Psi_{forbidden}$)}

The \textit{receiver-liability} in a unit-region $\chi$, at the ${\tau}^{th}$ snapshot of time, and in the frequency-band $\nu$  is defined as the receiver-liability at a sample point $\rho_0 \in \chi$ in the given unit-region 
\begin{equation}
\label{eq:urspfb}
{\Phi}(\chi, \tau, \nu) = \phi(\rho_0, \tau, \nu) . 
\end{equation}
The forbidden spectrum in a geographical area is quantified as the sum of receiver-liability in all the unit-regions across all the frequency bands of interest, in the $\hat{T}$ time-quanta.
\begin{equation}
\label{eq:agspfb}
\Psi_{forbidden} = \sum_{k=1}^{\hat{B}} \sum_{j=1}^{\hat{T}} \sum_{i=1}^{\hat{A}} {\Phi}(\chi_i, \tau_j, \nu_k) .
\end{equation}

\noindent
\textbf{Relationship between the Spectrum Consumption Spaces}
The spectrum consumption in a unit space $\chi$ is specified in terms of the spectrum-occupancy, spectrum-opportunity, and receiver-liability at a sample point $\rho_0 \in \chi$. Therefore,
\begin{equation}
{\Omega}(\chi, \tau, \nu) + {\Phi}(\chi, \tau, \nu)  + {\Gamma}(\chi, \tau, \nu)	= P_{MAX} .
\end{equation}
Summing over all the $\hat{A}$ unit-regions in the geographical-region, $\hat{B}$ frequency-bands, $\hat{T}$ unit-time quanta, we get following relation between the utilized spectrum, the forbidden spectrum, and the available spectrum.
\begin{equation}
\label{eq:p1e_sp_csm_reln}
\Psi_{utilized} + \Psi_{forbidden} + \Psi_{available}  = \Psi_{Total} .
\end{equation}

\subsection{Applying the Spectrum Consumption Methodology}

\subsubsection{Propagation Model Considerations}
So far we illustrated the \textit{base} spectrum consumption quantification methodology. While applying in practice, the above described spectrum consumption methodology requires choosing the appropriate propagation model that captures the path-loss and fading effects. 

\subsubsection{Unit Spectrum Space Granularity Considerations}
Spectrum consumption quantification in a geographical region, in a time interval, and within a spectral range is subject to error based on the rate at which spectrum is sampled in the time, space, and frequency dimensions. While applying the spectrum consumption methodology, the granularity of unit time, space, and frequency dimensions need to be chosen based on usage of the spectrum along these dimensions.

Besides this, the quantification methodology does not capture the spectrum use in the \textit{code} dimension of spectrum-access. Thus, it cannot be applied in the context of spectrum sharing with CDMA systems.
\section{Illustration: Spectrum Consumption Spaces}

\subsection{Spectrum Consumption at a Point}

\begin{figure}[htbp!]
\centering
{\includegraphics [width=0.46\textwidth, angle=0] {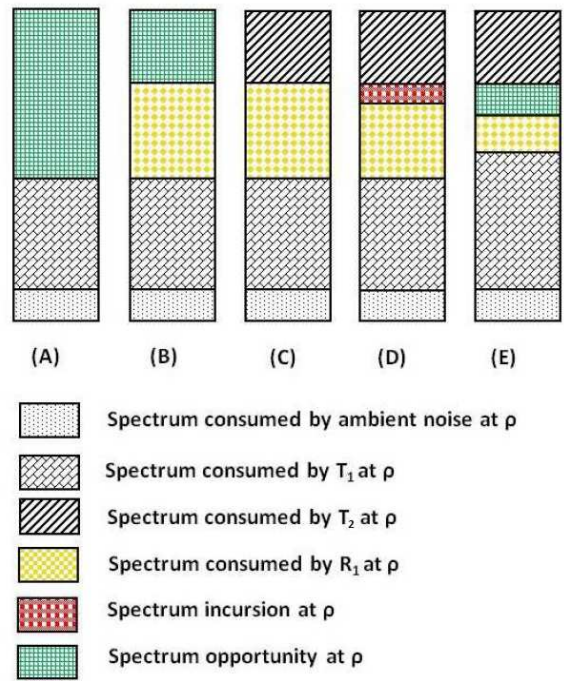}}
\caption{Quantification of spectrum consumption  at a point $\rho$ in a unit-region. The top of the bars represent the maximum power $P_{MAX}$ Watts and the bottom of the bars represent $P_{MIN}$ Watts. The figure shows five snapshots of spectrum consumption at $\rho$: (A) Spectrum consumed only by the transmitters and ambient noise (B) Spectrum consumed by transmitters, receivers, and ambient noise (C) Zero spectrum-opportunity (D) Negative spectrum-opportunity (spectrum-incursion)  (E) Reduced receiver-liability.}
\label{fig:p1e_sp_csm_pt}
\end{figure}

Figure \ref{fig:p1e_sp_csm_pt} shows five snapshots of spectrum consumption at $\rho$. In snapshot A, a certain volume of the spectrum is consumed by the ambient noise and the transmitter $T_1$. No receivers are assumed to be operating in this snapshot. Hence, no spectrum is consumed by the receivers and the remaining volume represents the spectrum-opportunity $\gamma(\rho)$ which is finite as $P_{MAX} < \infty$. 

In snapshot B, receivers are assumed to be operating. A certain volume of the available spectrum from snapshot (A) is consumed by the receivers. This volume represents the receiver-liability $\phi(\rho)$. We observe that the spectrum-opportunity, $\gamma(\rho)$, is still positive.

The snapshot C depicts the case when $\gamma(\rho)$ is zero. Here, a transmitter $T_2$ exploits all the spectrum-opportunity from the previous snapshot. If the power received from transmitter $T_2$ is increased, at least one of the receivers of $T_1$ will experience harmful interference. This is shown in snapshot D. 

In snapshot E, the transmit power of the transmitter $T_1$ is increased. (The spectrum consumed by transmitter $T_2$ is same as in snapshot C). This has effect of increasing the SINR for the receivers of $T_1$ and reducing the receiver-liability at $\rho$. 

\subsection{Spectrum Consumption in a Geographical Region}

Let us consider a region with 676 hexagonal unit regions with each side 100 m. Let the maximum RF-power at a point, $P_MAX$ in the unit region be 30 dBm i.e. 1 W. Let $P_MIN$ be a very low value, -200 dBm. Let us consider 6 MHz spectral range as unit bandwidth. Let us consider 10 second time period as unit time. In this case, the maximum spectrum consumption in the geographical region, in a 6 MHz spectral band, in 10 second time period is 676 $Wm^2$.

The propagation conditions are modeled by distance dependent path-loss with the path-loss exponent as 3.5. It should be noted that the base quantification methodology is independent of the propagation model applied while spectrum consumption.

\subsubsection{Spectrum Consumed by a Transmitter}
\begin{figure}[htbp]
\centering
{\includegraphics [width=0.48\textwidth, angle=0] {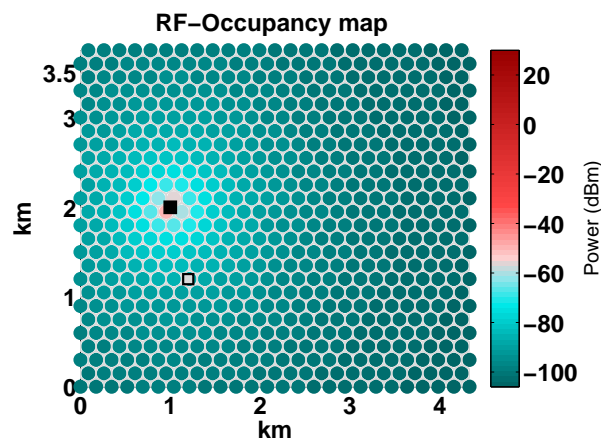}}
\caption{The Spectrum-occupancy map shows spatial distribution of the discretized spectrum occupancy, that is the spectrum consumed by a transmitter. } 
\label{fig:L201}
\end{figure}

Figure~\ref{fig:L201} illustrates the spectrum consumed by a transmitter device given by (\ref{eq:urspoc}). The transmitter is located at (1000, 2000) and is exercising omnidirectional transmission with transmit power 15 dBm. The spectrum consumed by the transmitter is $1.8$ x $10^{-8}$ $Wm^2$ ($2.7$ x $10^{-9}$ \%). 

\subsubsection{Spectrum Consumed by a Receiver}

\begin{figure}[htbp]
\centering
{\includegraphics [width=0.48\textwidth, angle=0] {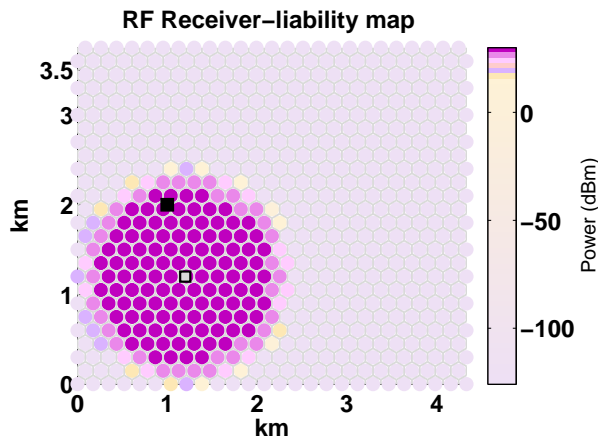}}
\caption{The receiver-liability map shows the spatial distribution of receiver-liability, that is the RF-power consumed by a receiver at a point in space.} 
\label{fig:L202_RL}
\end{figure}

Figure~\ref{fig:L202_RL} illustrates the spectrum consumed by a receiver device given by (\ref{eq:urspfb}). The receiver is located at (1200, 1200) and is exercising omni-directional reception with minimum SINR of 6 dB and the actual experienced SINR at the reciever is 33 dB. We note that as the distance from a receiver increases, a cochannel transmitter can exercise higher transmission power. Thus, the liability for ensuring non-interference to the receiver goes down with the distance from the receiver. The spectrum consumed by the receiver is $112.4$ $Wm^2$ (16.6\%).


\subsubsection{Available Spectrum in a Geographical Region}
Figure~\ref{fig:L202_RL} illustrates the spectrum not consumed by the transmitters and receivers, that is the available spectrum in the geographical region. The spectrum-opportunity map shows spatial distribution of spectrum-opportunity given by (\ref{eq:urspop}). The topology described is same as described in Figure \ref{fig:L202_RL} with the transmitter is located at (1000, 2000) and the receiver is located at (1200, 1200). The available spectrum is $563.6$ $Wm^2$ (83.4\%).
\begin{figure}[htbp]
\centering
{\includegraphics [width=0.48\textwidth, angle=0] {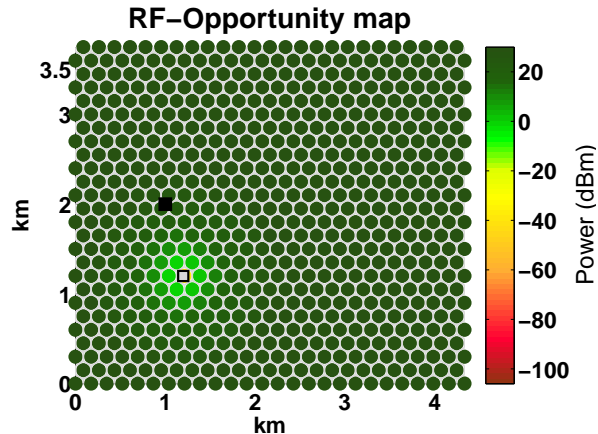}}
\caption{The spectrum-opportunity map shows the spatial distribution of spectrum-opportunity, that the additional RF-power that can be tolerated by the cochannel receivers.}
\label{fig:L202_OP}
\end{figure}


\noindent
\subsection{Illustration of the Error due to Discretization}
\begin{figure}[htbp!]
\centering
{\includegraphics [width=0.32\textwidth, angle=0] {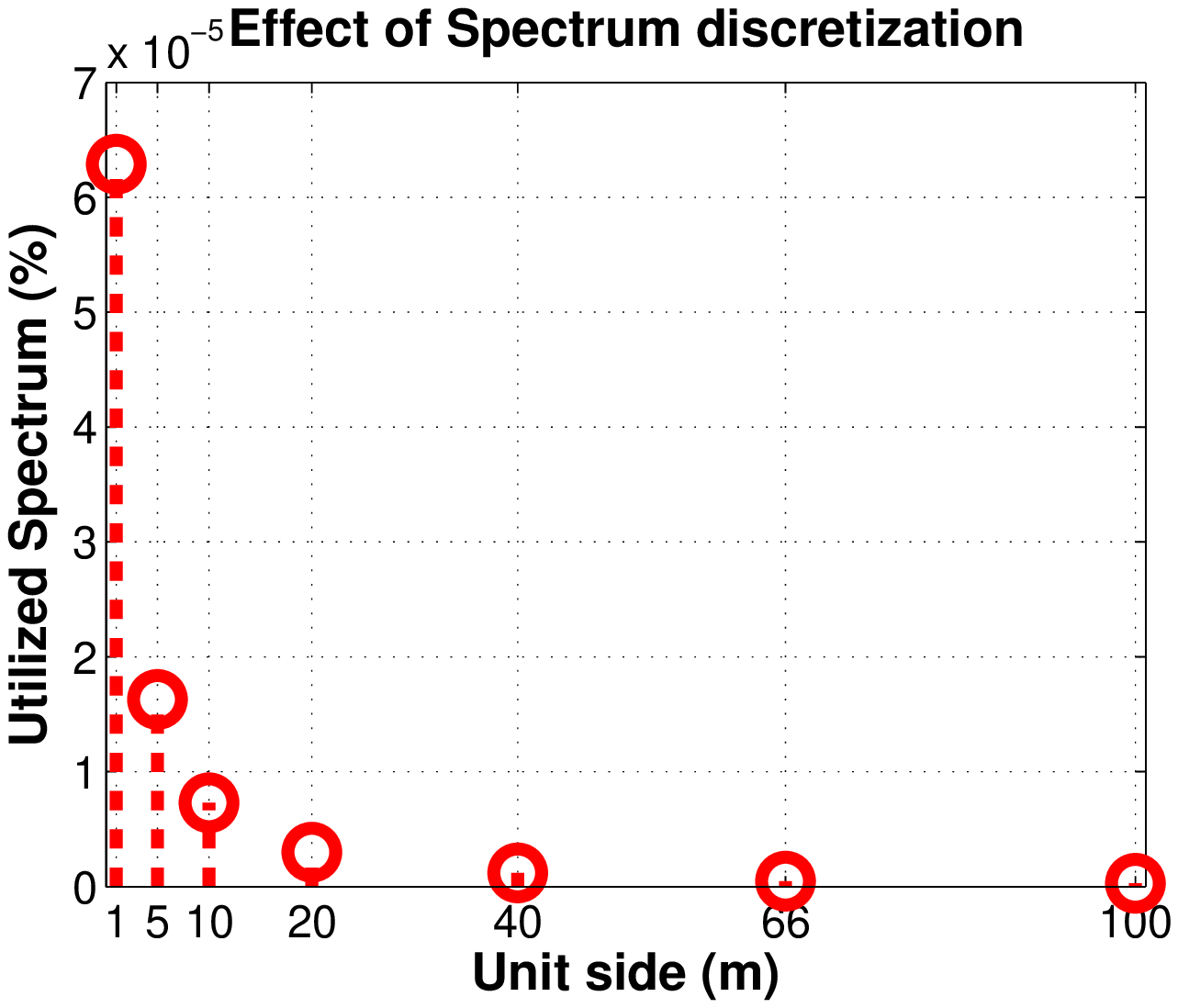}}
{\includegraphics [width=0.32\textwidth, angle=0] {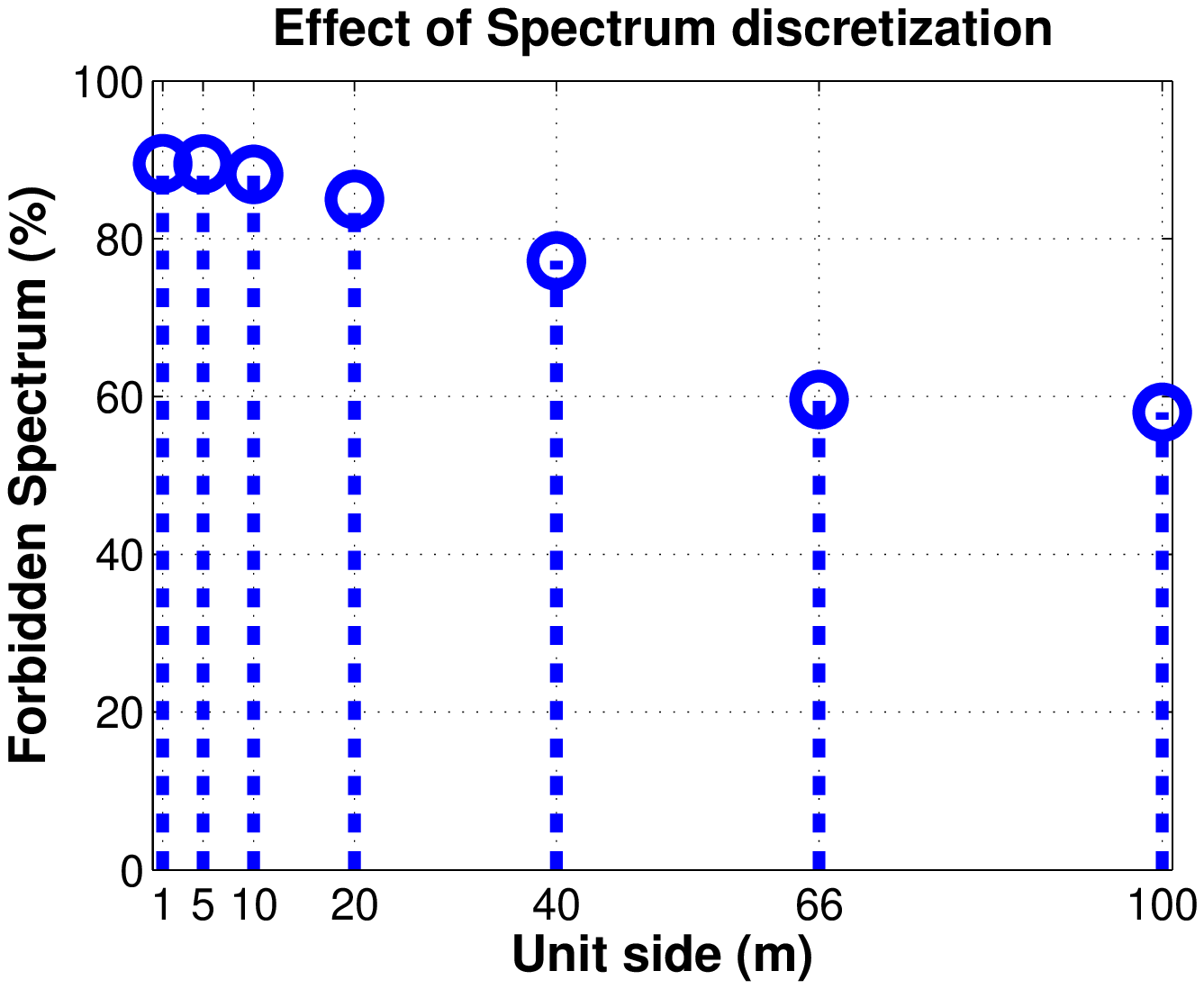}}
{\includegraphics [width=0.32\textwidth, angle=0] {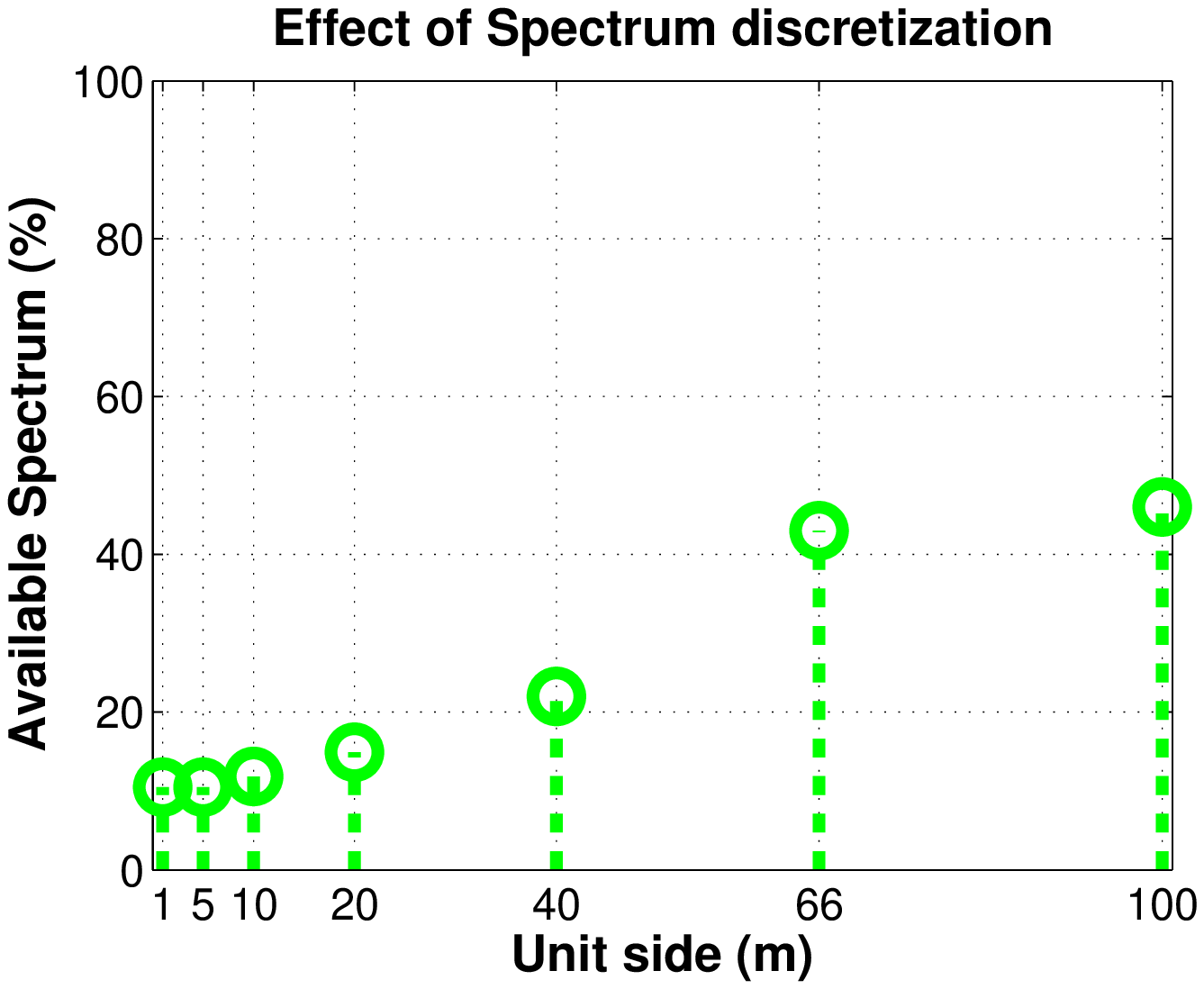}}
\caption{The plots capture the effect of spectrum discretization on the quantification of the spectrum consumption spaces. The $x$ axis shows the length of unit-side of the hexagonal unit regions. The spectrum consumption (spectrum-occupancy, spectrum-opportunity, and receiver-liability) in a unit-region is governed by spectrum consumption at the sample point. Thus, we observe that the quantification of the spectrum consumption spaces vary depending on unit-side length which represents the spatial sampling rate.}
\label{fig:L207}
\end{figure}

The rate at which spectrum-occupancy, receiver-liability, and spectrum-opportunity are sampled in the time, space, and frequency dimensions determine the error in the quantification of the utilized, forbidden, and available spectrum consumption spaces respectively. Figure \ref{fig:L207} illustrates the error in quantification of spectrum consumption spaces with the spatial sampling rate. When the side of the unit-region is 100 m, the available spectrum is much higher as compared the available spectrum when the side of the unit-region is 1 m.  Note that the receiver distance is kept maximum from the UR-centers in order to maximize the spectrum availability and consequently the error due to discretization. Thus, when the unit-side is 1 m , the sampling point is very close to the actual receiver location and the spectrum-opportunity is accurately quantified. This difference in the quantification of the available spectrum illustrates that sampling rate should be chosen based on the \textit{access-density} or the \textit{propagation environment characteristics}.   

\section{Transition from Continuous Spectrum Space to Discretized Spectrum Space}

Discretization of the spectrum space transforms the spectrum hole detection approach for spectrum sensing which infers spectrum access opportunities as '\textit{white}' or '\textit{black}' to \textit{quantified spectrum harvesting} approach wherein the spectrum consumption spaces are estimated. In this regard, we note that the error in spectrum hole detection due to the \textit{false positives} are translated into \textit{lost available spectrum space}. On the other hand, missed detection of a signal would be translated into \textit{potentially degraded spectrum space}. Similarly, in case of spectrum allocation, when a spectrum allocation mechanism does not exploit all the available spectrum identified by a spectrum harvesting mechanism,  it implies a \textit{unexploited spectrum space}. If the spectrum allocation mechanism allocates spectrum in such a way that causes harmful interference to cochannel receivers, it suggests a \textit{degraded spectrum space}. Thus, by quantifying above mentioned spectrum consumption spaces, the performance of spectrum harvesting and spectrum exploitation techniques could be quantified and analyzed.

Spectrum-space discretization implies the shared spectrum could be \textit{quantified} instead of implied based on the spectrum-access parameters of the transceiver devices. A dynamic spectrum-access policy could be defined to assign quantified spectrum resource to each of the spectrum-access requests. 

The dynamicities and uncertainties pertaining to the RF environment as well as the spectrum-access conditions are the key challenges to realizing the potential of dynamic spectrum sharing. In this regard, the proposed spectrum-discretization approach facilitates application of machine-learning techniques to spectrum consumption management and is attractive for bringing in learning and adaptation to spectrum harvesting and exploitation mechanisms.  

Tables I and II \mycomment{\ref{table:SHSCS} and \ref{table:SXSCS}} describe spectrum consumption spaces in case of spectrum recovery and spectrum exploitation functions respectively. 

\begin{table}[h!b!p!]
\label{table:SHSCS}
\caption{Spectrum Consumption Spaces in case of Analysis of Spectrum Recovery}
\centering
\begin{tabular}{lp{5cm}p{5cm}}
\hline
Space & What it represents & Notes\\
\hline
Available Spectrum  & It represents the actually available spectrum in case of a spectrum access scenario specified RF environment conditions & Since the \textit{absolute} available spectrum has been quantified, it serves as the reference in the analysis or comparison of the spectrum recovery techniques.\\
Potentially Degraded Spectrum & [TUH] It represents the portion of non available spectrum that has been erroneously treated as available spectrum. This is the recovered spectrum and not actually exploited. Thus, the spectrum space suggests potentially harmful interference at the receivers.& An example of a case leading to potentially degraded spectrum space is \textit{missed detection} event at the detector identifying presence of the transmitters.\\
Potentially Degraded Spectrum & It represents the portion of non available spectrum that has been erroneously treated as available spectrum. This is the recovered spectrum and not actually exploited. Thus, the spectrum space suggests potentially harmful interference at the receivers.& An example of a case leading to potentially degraded spectrum space is \textit{missed detection} event at the detector identifying presence of the transmitters.\\
\hline
\end{tabular}
\end{table}

\begin{table}[h!b!p!]
\label{table:SXSCS}
\caption{Spectrum Consumption Spaces in case of Analysis of Spectrum Exploitation}
\centering
\begin{tabular}{lp{5cm}p{5cm}}
\hline
Space & What it represents & Notes\\
\hline
Available Spectrum  & [TUH] It represents the actually available spectrum in case of a spectrum access scenario specified RF environment conditions & Since the \textit{absolute} available spectrum has been quantified, it serves as the reference in the analysis or comparison of the spectrum recovery techniques.\\
Potentially Degraded Spectrum & [TUH] It represents the portion of non available spectrum that has been erroneously treated as available spectrum. This is the recovered spectrum and not actually exploited. Thus, the spectrum space suggests potentially harmful interference at the receivers.& An example of a case leading to potentially degraded spectrum space is \textit{missed detection} event at the detector identifying presence of the transmitters.\\
Potentially Degraded Spectrum & [TUH] It represents the portion of non available spectrum that has been erroneously treated as available spectrum. This is the recovered spectrum and not actually exploited. Thus, the spectrum space suggests potentially harmful interference at the receivers.& An example of a case leading to potentially degraded spectrum space is \textit{missed detection} event at the detector identifying presence of the transmitters.\\
\hline
\end{tabular}
\end{table}

\section{Identifying Opportunities for Efficient Spectrum Sharing}
Discretization of the spectrum-space and the spectrum consumption quantification methodology can facilitate analysis of the spectrum management functions. In this section, we discuss how it can be used to understand the weaknesses and improve performance of secondary spectrum access. 
 
\subsection{Secondary Spectrum Access Scenario of Spectrum Sharing}
One of the popular use cases for spectrum sharing is exploiting the underutilized spectrum on secondary basis. While this seems to be a reasonable solution to the spectrum scarcity problem, the owners of the spectrum are concerned about disruption or degradation of their service due to harmful interference from the secondary users. Also, the performance estimation studies of secondary access have revealed that the amount of the secondary spectrum that could be available is very limited to meet the increasing demand for RF spectrum \cite{berk_wsc, osa_feasib}.
 
To address the incumbents' concerns, there is a need of a shared spectrum access paradigm that can enable controlling and enforcing spectrum-access parameters. In this regard, the proposed spectrum consumption quantification methodology enables us to quantify the spectrum consumed by a spectrum-access request and facilitates the \textbf{\textit{enforcement of spectrum access policy}} in the discretized spectrum space. Thus, non-harmful interference to the licensed (as well as secondary) receivers could be ensured. 

From the secondary-user perspective, the availability of the secondary spectrum, throughput, and quality of service need to be ensured. In this regard, we need the ability to characterize the performance of recovery and exploitation of the underutilized spectrum. Often, the performance of the spectrum recovery mechanisms is captured in terms of Receiver Operating Characteristics (ROC) curve; however it does not provide with the absolute spatial, spectral, or temporal spectrum recovered. Similarly, The average throughput and BER metrics cannot be used in quantifying the exploitation of the recovered secondary spectrum. The proposed spectrum consumption quantification methodology can be applied to quantify the available spectrum rendered unexploitable due to missed detection or conservative spectrum access policy. Thus, it is possible to investigate why the existing approaches are not able to \textit{recover} the underutilized spectrum. We illustrate performance analysis of spectrum recovery in case of OSA mechanism in Appendix B. By incorporating the knowledge of the spectrum-access parameters of the cochannel transceivers and estimation of the RF environment conditions, it is possible to estimate spectrum consumption and identify the underutilized spectrum in real time. Furthermore, it is also possible to investigate how efficiently the underutilized spectrum gets exploited by various secondary spectrum access approaches. In Appendix C, we compare performance of various SAMs in terms of their ability to effectively exploit the available spectrum. In Appendix D, we describe the spectrum consumption estimation approach.





\section{Quantified Dynamic Spectrum Access Paradigm}

In this section, we apply the spectrum consumption quantification model for provisioning quantified spectrum access to multiple heterogeneous networks and manage the spectrum consumption in a geographical region. 

Under quantified dynamic spectrum access paradigm, the spectrum-consumption is quantified in terms of unit-regions in the spatial dimension, spectral-band in the frequency dimension, and time-quanta in the temporal dimension. A spectrum-access footprint represents the spectrum resource quantified in the space, time and frequency dimensions. 

A spectrum-access policy represents the spectrum access attributes for a spectrum resource, given the transmitter and receiver positions, propagation medium, and expected link quality. 

A service provider, a spectrum broker, or a proprietary network can request for spectrum-access to Spectrum-access Policy Infrastructure (SPI). SPI assigns and enforces allocated spectrum-access-policies. Spectrum management infrastructure (SMI) schedules spectrum-access requests and defines spectrum-access parameters in order to manage spectrum consumption. Spectrum Analysis Infrastructure (SAI) estimates the spectrum-access footprints and the available spectrum in real time using the RF environment data acquired by Spectrum Sensing Infrastructure (SSI).

\subsection{Estimating the Available Spectrum}
Knowing how much spectrum is available requires the knowledge or estimation of the transceiver parameters. It also requires the estimation of propagation medium parameters in order to estimate spectrum consumption by transmitters and receivers. In this regard, we resort to an external RF-sensor network that helps to address the underlying signal-processing subproblems, estimates the spectrum-access parameters, estimates the spectrum consumption spaces for each of RF-sensors, and finally fuses the estimates spectrum-consumption spaces. Appendix D provides more details on the spectrum consumption estimation approach.

Using this spectrum consumption estimation approach, SSI provides the spectrum-opportunity estimates for each of the frequency-bands, for each of the unit-regions in the geographical region. Thus, spectrum-space discretization and the quantification methodology enables us to identify the maximum transmit-power for a potential transmitter ahead of time! Therefore, it very much suitable for applying it to real-time dynamic spectrum access.

\subsection{Defining a Dynamic Spectrum-access Policy}
We note that the spectrum consumption by transmitters is dependent on
\begin{enumerate}
	\item the actual (as against the maximum) transmit power.
	\item the antenna directionality employed during transmission.
	\item the location of the transmitter. In case of transmitter mobility, the spectrum-access policy needs to be periodically updated.
\end{enumerate}
\textit{In order to \textbf{control} the spectrum consumption by \textbf{transmitters}, these transmitter parameters should be part of the spectrum-access footprint.}

Similarly, the spectrum consumption by receivers is dependent on
\begin{enumerate}
	\item the minimum SINR required for successful reception.
	\item the antenna directionality employed during reception.
	\item the location of the receiver. In case of receiver mobility, the spectrum-access policy needs to be periodically updated.
\end{enumerate}
\textit{In order to \textbf{control} the spectrum consumption by the \textbf{receivers}, these receiver parameters should be part of the spectrum-access footprint.}

SMI needs to schedule the spectrum-access requests and assign \textit{quantified} spectrum-access footprint using the above mentioned transceiver parameters, a frequency band(s) and the temporal boundaries. A SAM defines the interference management approach and plays a crucial role in defining how efficiently spectrum can be shared by multiple spatially overlapping heterogeneous wireless networks. In appendices B and C, we compare performance of various SAMs and investigate the SAM design choices that can maximize the exploitation of spectrum. 

Quantification of spectrum consumption spaces helps us in addressing the joint scheduling and power allocation problem. This problem is NP-hard. In appendix C, we describe network-spectrum-consumption based coexistence (NSC-CX) SAM that favors spectrum-access requests with lower network spectrum consumption and also ensures coexistence with cochannel (primary and secondary) networks.
\subsection{Enforcement of Spectrum-access Policy}
Quantified spectrum-access policy enables us to regulate the spectrum access by detecting violations of the assigned spectrum access policy. When transceiver devices do not conform to the spectrum-access constraints imposed by the spectrum-access policy, it may lead to degradation of link-throughput and service-disruption for the cochannel networks. 

\begin{figure}[htbp!]
\centering
{\includegraphics [width=0.66\textwidth, angle=0] {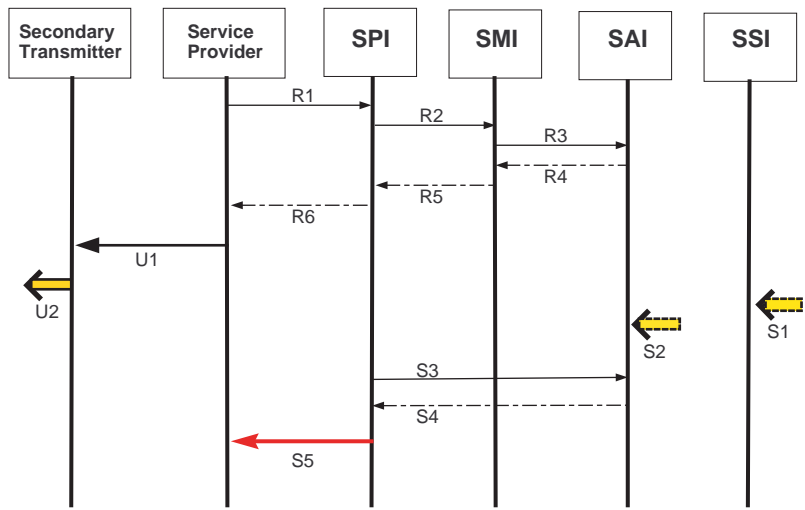}}
\caption{Illustration of defining and enforcing a spectrum-access policy. A service-provider requests a spectrum-access footprint from SPI along with the information about position and capabilities of the transceivers (Arrows R1-R6). A service-provider assigns a partial time to one of the secondary transmitters (Arrow U1). The secondary transmitter fails to conform to the assigned quantified spectrum-access policy (Arrow U2). This scenario is detected by transmitter footprint estimation (Arrows S1-S4) and regulatory action is taken (Arrow S5).}
\label{fig:qdsa_use_case}
\end{figure}

Conformance to spectrum-access policy can be enforced via \textit{estimation} of the spectrum consumption by the transmitters \cite{oms4_sce}. Thus, the RF-sensor network used for estimating the available spectrum can be used for estimating the transmit-power of the transmitters. Thus, the fusion-center generates estimates spectrum consumption spaces for all the cochannel transmitters. SPI can determine if the actual spectrum consumed by a transmitter is unacceptable, and initiate a regulatory action. We illustrate this scenario in Figure \ref{fig:qdsa_use_case}.
Here, we note that when receivers do not comply with the receiver parameters specified in the spectrum-access footprint, the receiver spectrum consumption gets incorrectly quantified and it may result into harmful interference experienced at the receiver. As non-conformance by receivers does not cause any harm to other services, no regulatory action is needed. 

%

\section{Conclusions and Future Research Avenues}
Discretization of an analog signal brings in enormous advantages and opens up countless opportunities. In this paper, we have applied the same concept to spectrum-space dimensions. We discretized the spectrum consumption spaces and defined a methodology to quantify, estimate, analyze, and optimize the various spectrum consumption spaces in the generalized spectrum usage scenario.  

Spectrum-space discretization and the spectrum consumption quantification methodology help to address several challenges in the recovery and exploitation of the underutilized spectrum. First, it facilitates quantifying the spectrum recovered by a spectrum recovery mechanism. This helps to understand the weaknesses of a given spectrum recovery mechanism. It allows comparing performance of multiple spectrum recovery mechanisms. Similarly, it enables quantifying, analyzing, and comparing the performance of the spectrum exploitation mechanisms. The quantification of spectrum consumption spaces brings in a quantified dynamic spectrum access paradigm and can facilitate real-time fine grained dynamic spectrum sharing. The spectrum-discretization approach facilitates application of machine-learning techniques to spectrum consumption management and is attractive for bringing in learning and adaptation to spectrum harvesting and exploitation mechanisms.

We note that the discretized spectrum-space concept and the spectrum consumption quantification methodology introduced in this paper are only the initial steps towards spectrum consumption management techniques necessary for spectrum sharing paradigm. Much research is needed in the analysis and optimization of spectrum management functions in order to realize the promise held by dynamic spectrum sharing paradigm.


\appendices


\section{Performance Analysis of Spectrum Recovery}
In this appendix, we apply spectrum-space discretization to quantify the recovered spectrum based on the constraints imposed by a spectrum access mechanism (SAM). This provides us insights on how to improve spectrum recovery.

Here, we model the spectrum access constraints under Opportunistic Spectrum Access (OSA) model that allows secondary access to the spectrum only when the primary user (PU) is not operating or when the secondary user (SU) is outside the PU operating range. The PU receivers may be passive and SU receivers are required to detect the presence of the PU signal and assume that the PU receivers are at their worst-case positions.

\mycomment{
The constraint on the minimum sensitivity at the SUs play an important role in spectrum recovery. Table \ref{table:OSASensingRadius} shows the effective sensing radius or detection radius $R_D$ for various sensitivity, path-loss exponent (PLE), and PU transmit power combinations.We observe that \textsl{when the required SU sensitivity is -90 dBm and the assumed PLE is 2.5, SU is not able to exercise secondary access even when PU transmitter is 63 km away from the SU transmitter.} 
\begin{table}[h!b!p!]
\label{table:OSASensingRadius}
\caption{Sensing radius implied by the SU sensitivity constraint}
\centering
\begin{tabular}{llll}
\hline
Sensitivity (dBm) & Max PU $T_p$(dBm) & PLE & Sensing Radius (m)\\
\hline
-60  & 30 & 3.5 & 372\\
-90  & 30 & 3.5 & 2682\\
-120  & 30 & 3.5 & 19307\\
-60  & 30 & 2.8 & 1638\\
-90  & 30 & 2.8 & 19307\\
-60  & 30 & 2.5 & 3981\\
-90  & 30 & 2.5 & 63096\\
\hline
\end{tabular}
\end{table}
}

\begin{figure}[htbp!]
\centering
{\includegraphics [width=0.45\textwidth, angle=0] {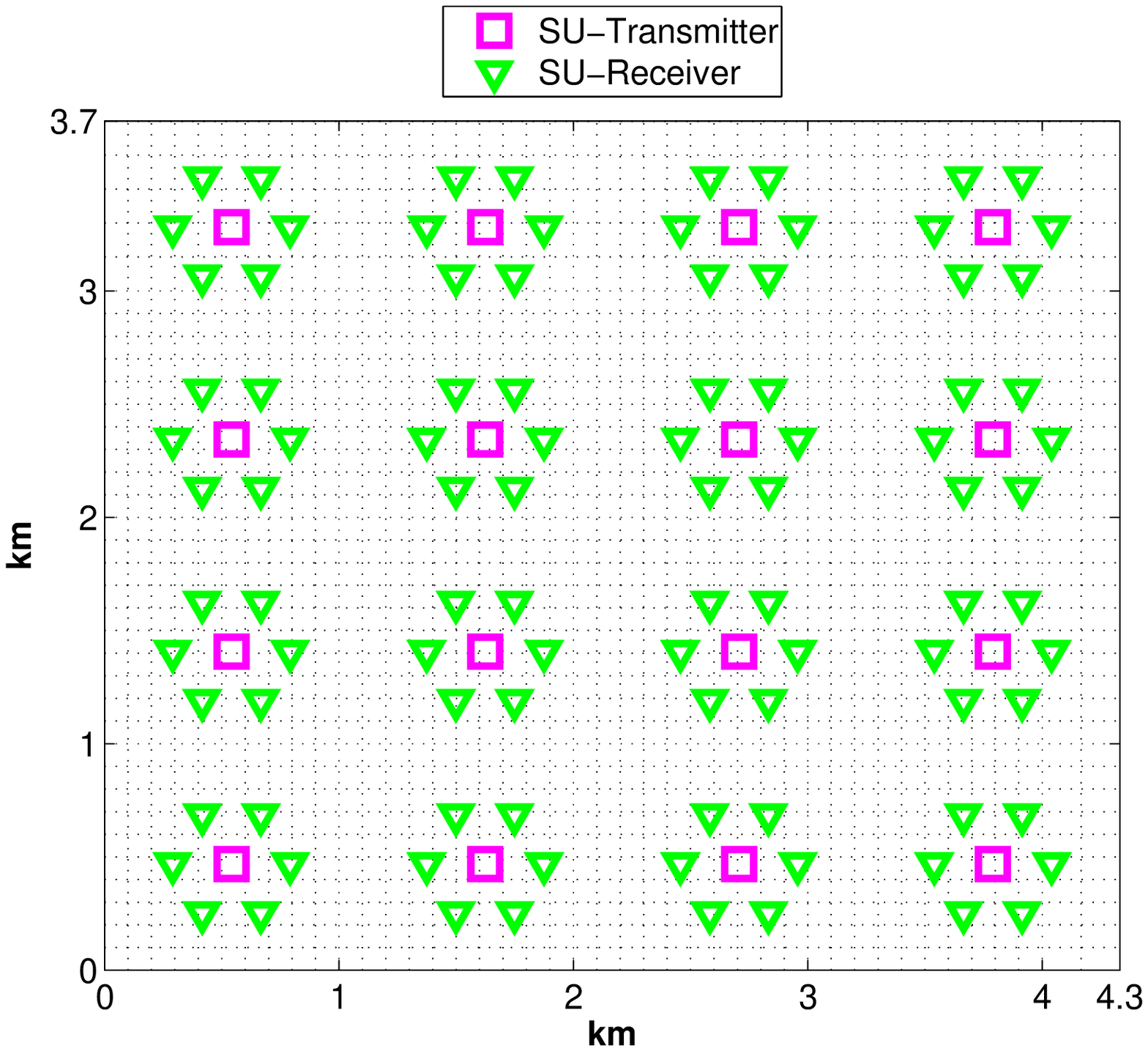}}
{\includegraphics [width=0.45\textwidth, angle=0] {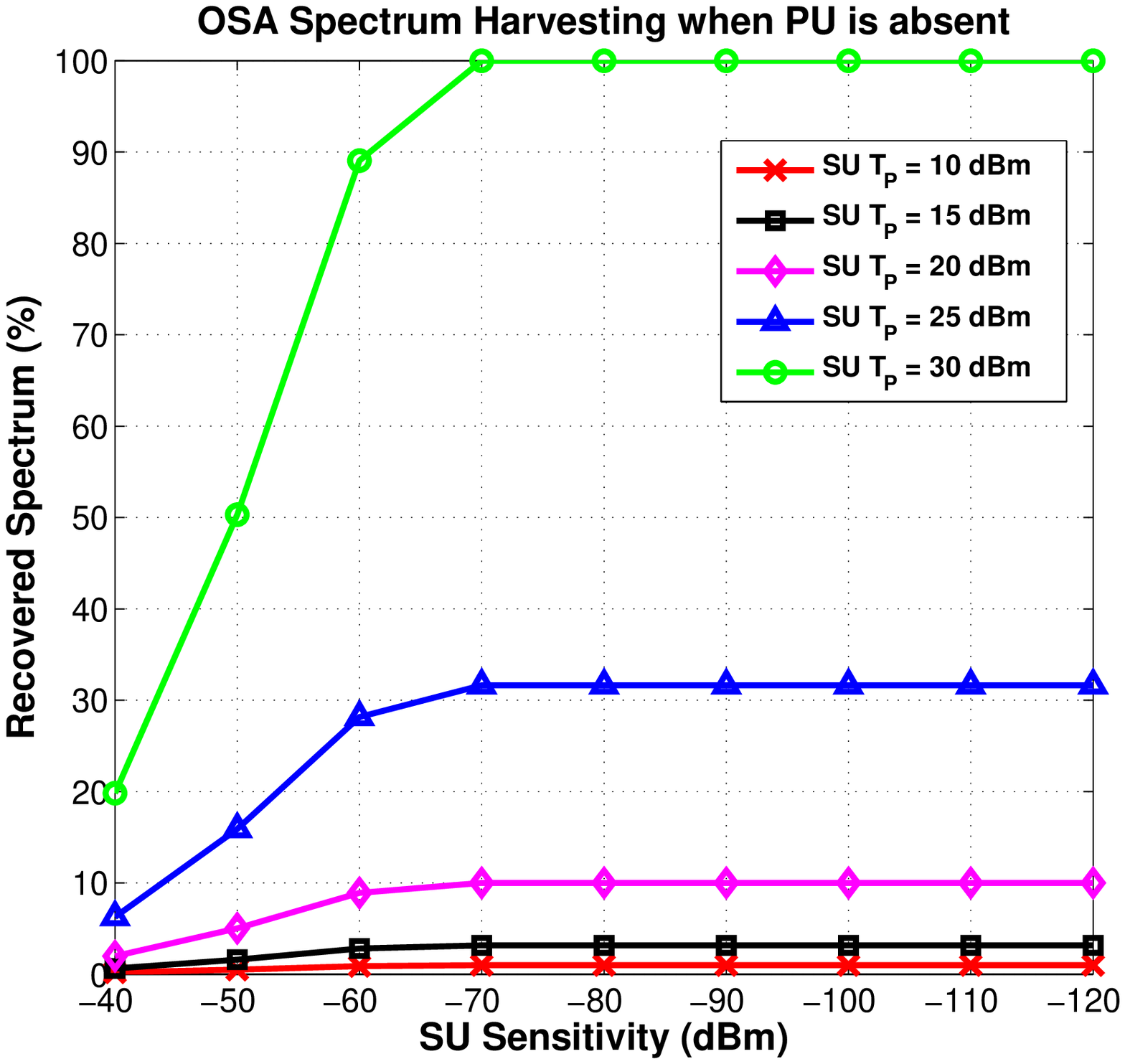}}
\caption{The performance of spectrum recovery when PU is absent. The left plot shows 16 SU networks with 112 transceivers. Per sensitivity and the maximum transmit power constraint specified by the OSA policy, the 112 SU transceivers recover the spectrum. The right plot shows that as the sensitivity increases, the detection range increases, and more spectrum is recovered. The lower limits on the maximum transmit power result into poor recovery of the spectrum.} 
\label{fig:L301_PU_ABSENT}
\end{figure}

We argue that the constraint on the maximum SU transmit-power in order to avoid harmful interference at PU receivers implies significant loss of the available spectrum. We setup an experiment to recover the available spectrum in a $4.3$ km x $3.7$ km geographical region when PU is \textit{not} operating. Figure \ref{fig:L301_PU_ABSENT} shows the performance of spectrum recovery. The left plot shows the topology with 112 SU transceivers sensing the channel to determine the presence of the primary transmitter and inferring if the secondary access is permitted. The right plot shows that the amount of the available spectrum that could be used for secondary access increases with the maximum secondary transmit power for specified sensitivity. A key observation from the spectrum harvesting performance when PU is absent is that the amount of spectrum harvested by 112 SUs with -120 dBm sensitivity in a $4.3$ km x $3.7$ km geographical region \textbf{is close to 0 \%} when the maximum transmit power for SUs is 10 dBm. As the limit on the maximum SU transmit power is raised to 30 dBm, more and more of the underutilized spectrum is harvested. This is due to the difference between the maximum transmit power in a unit spectrum space and the limit on maximum secondary transmit power. This difference results in significant loss of available spectrum. Based on the higher spectrum recovery performance with the higher values of the maximum permitted SU transmit power from Fig.\ref{fig:L301_PU_ABSENT}, we observe the need to \textit{dynamically compute the secondary transmit power based on the interference margin at the receivers}.

The amount of spectrum recovered \textit{when PU is present} is found to be \textit{close to 0\%} \cite{oms2_sca}. This is primarily due to the impact of the constraint on minimum SU sensitivity. We observe that when the required SU sensitivity is -90 dBm and the assumed mean path-loss exponent is 2.5, a secondary user is not able to exercise secondary access even when it is 63 km away from the primary transmitter. Please refer to \cite{oms2_sca} for more details.


\section{Performance Comparison of Spectrum Access Mechanisms}
Spectrum-space discretization enables us to quantify the exploited spectrum space, degraded spectrum space, and unexploited or available spectrum space. Here, we investigate the impact of various interference management choices adopted by spectrum access mechanisms (SAMs) and compare performance of various SAMs.

\subsection{Metrics}
From a spectrum-provider's or an incumbent's perspective, the preferred SAM minimizes the consumption of the this resource while maximizing the number of scheduled spectrum-access requests and minimizes the number of the harmfully interfered receivers. We compare the performance of SAMs in terms of
\begin{itemize}
	\item the number of scheduled spectrum-access requests
  \item the number of harmfully interfered receivers 
  \item the percentage of the spectrum exploited
	\item the percentage of the spectrum that remained unexploited (the available spectrum)  
\end{itemize}

\subsection{Setup}
We consider a geographical region with a primary transmitter at the center. The secondary users (SU) do not have knowledge of the positions of the primary user (PU) receivers which are assumed to be at the worst-case positions. The SU networks are scattered in the geographical region with the transmitter positions and the range of the SU networks being random variables with uniform distribution. We use distance dependent path-loss model with the mean path-loss exponent is 3.5. 
\begin{itemize}
	\item The minimum desired SINR for the worst-case PU receivers is 10 dB and the SINR experienced is 20 dB. The minimum desired SINR for secondary receivers is 3 dB.
	\item The range of the PU networks is 500 m and the range of the SU networks is 100 m. 
	\item The PU and SU receivers are employing directional reception. The SU transmitters are also assumed to employ directional transmissions. The antenna beamwidth for directional transmission and reception is assumed to be $60^\circ$. 
\end{itemize}



\subsection{The SAM Candidates}

Per the \textbf{Underlay} spectrum access mechanism \cite{dsa_survey}, the secondary users (SUs) exploit the spectrum with a very low transmit power in order to not cause severe interference at the primary user (PU) receivers. We consider the secondary transmit power to be 30 dB above the thermal noise power (-106 dBm for 6 MHz band). The underlay approach does not require to check whether the primary network is active at this time and location.

The second approach to spectrum access is the \textbf{Overlay} approach which requires secondary user devices to confirm that the primary transmitter signal is not present before it can access the spectrum with constrained transmit power \cite{dsa_survey}. The key concern with this approach is the sensitivity required for PU detection maps to a large spatial range \cite{oms2_sca}. Thus, in most of the spatial locations, the spectrum could not be exercised when the primary network is active.  

The third approach we investigate is an enhancement to the previous overlay approach which employs time-invariant constrained transmit power. It uses \textit{dynamic} transmit power in order to ensure high SINR for its receivers and protect those receivers from cochannel interference from other SU networks. \mycomment{It however cannot ensure non-harmful interference to the primary network users that are just beyond the sensing range of the secondary user network due to interference aggregation effect.} We term this approach as Secondary Throughput-oriented OVerlay SAM \textbf{(STOV DSA)}.

The final SAM approach assumes knowledge of the locations of the primary receivers and thus can correctly infer the interference margin imposed by the primary receivers. However, when multiple secondary networks are exercising secondary access with transmit power implied by the interference margin, the primary receiver may still experience harmful aggregate interference. To avoid this interference aggregation scenario, a \textit{guard margin} is used when inferring the transmit power for the secondary network.  We term this approach Secondary Throughput and Primary Protection oriented OVerlay SAM \textbf{(STPPOV DSA)}.

Table III \mycomment{\ref{table:SAMCandSum}} shows the summary of spectrum access strategies of the SAM candidates. 
\begin{table}[htbp!]
\label{table:SAMCandSum}
\caption{Summary of the SAM Properties}
\centering
\begin{tabular}{lll}
\hline
SAM & Potential Interference & Constraint on SU transmit-power\\
\hline
Underlay DSA  & to PU and SU & Fixed and low\\
Overlay DSA & to PU and SU & Fixed and high\\
STOV DSA  & to PU and SU & Dynamic and high\\
STPPOV DSA  & to SU only & Dynamic and high\\
\hline
\end{tabular}
\end{table}

\subsection{Observations}
\begin{figure}[htbp!]
\centering
{\includegraphics [width=0.39\textwidth, angle=0] {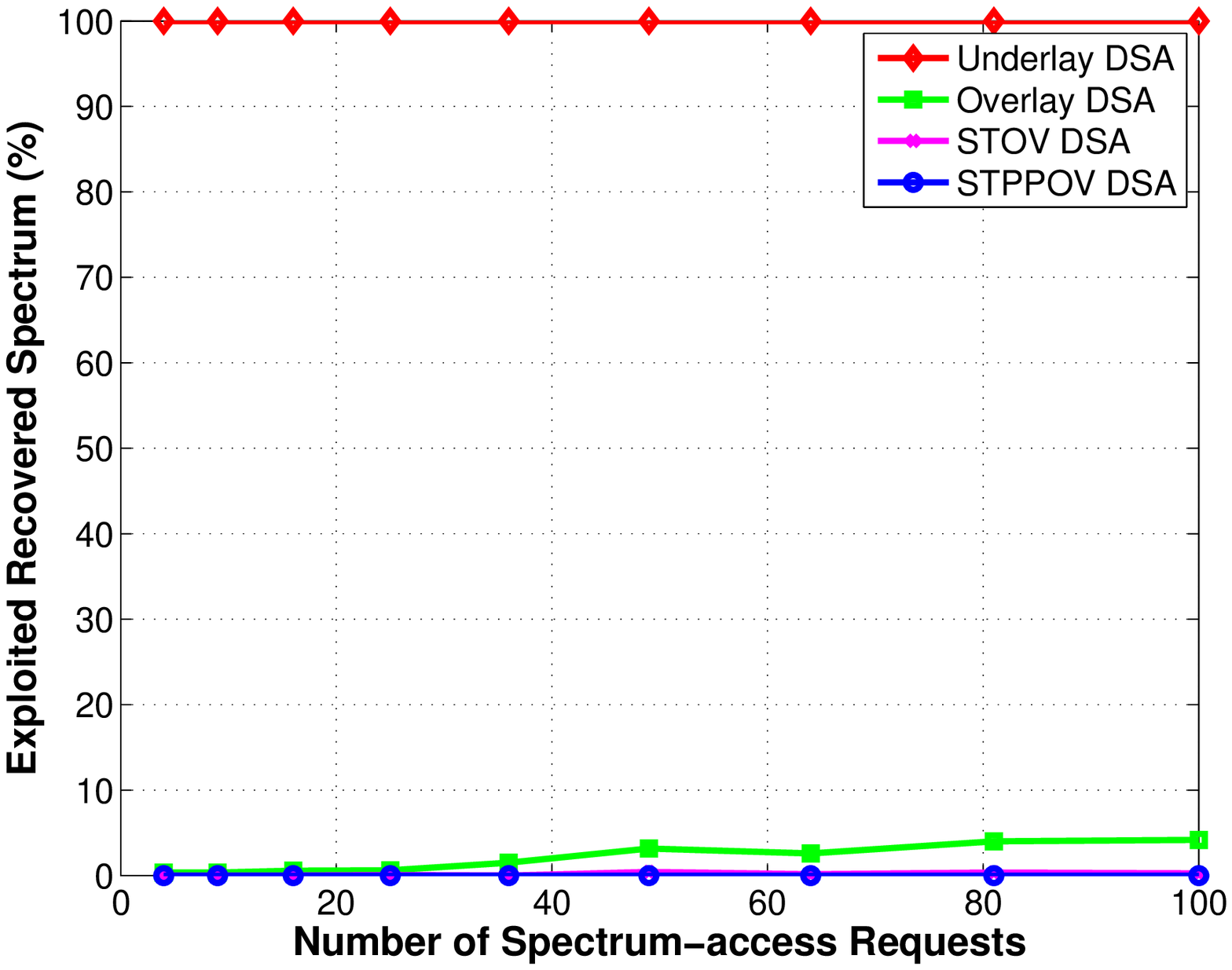}}
{\includegraphics [width=0.39\textwidth, angle=0] {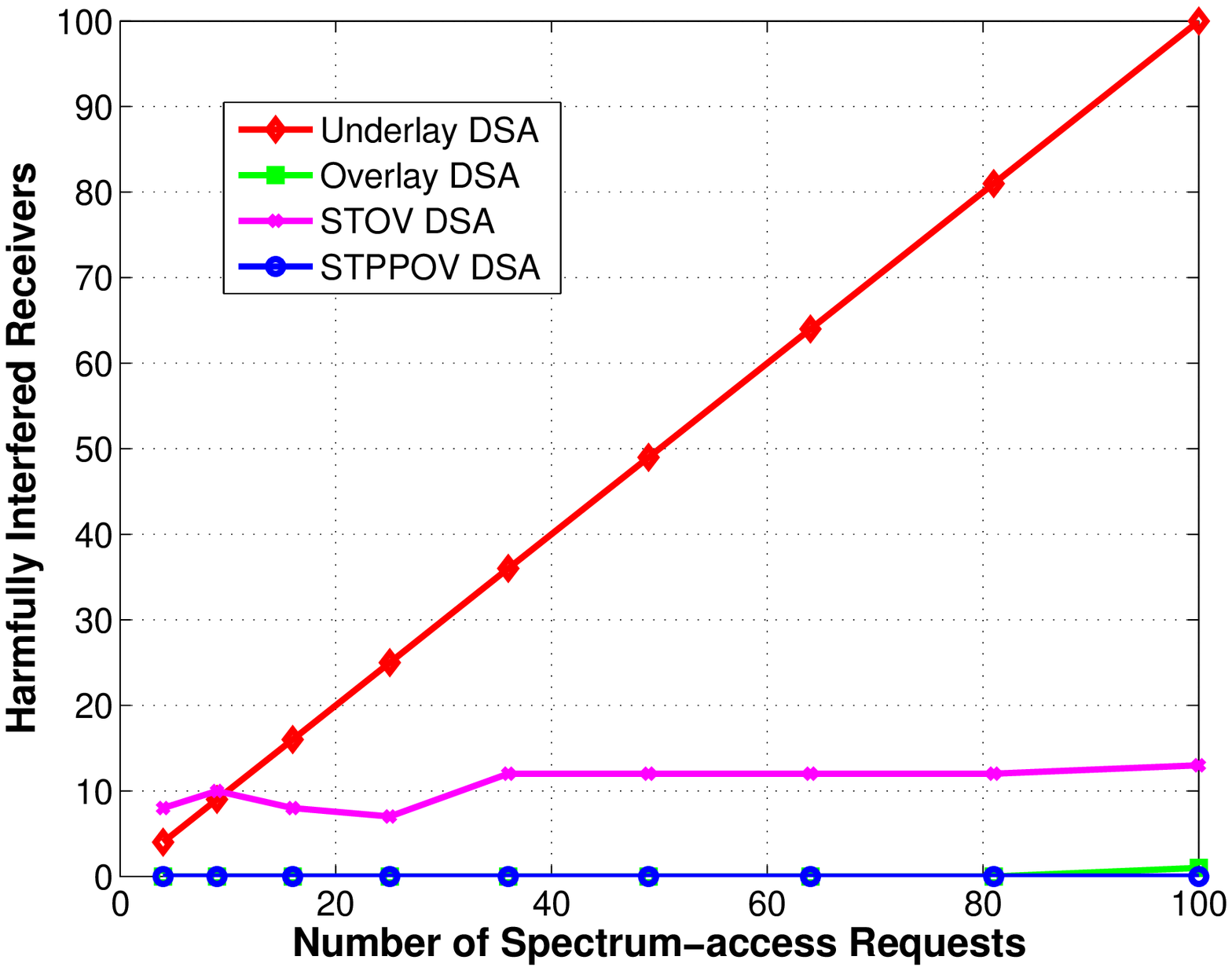}}
{\includegraphics [width=0.39\textwidth, angle=0] {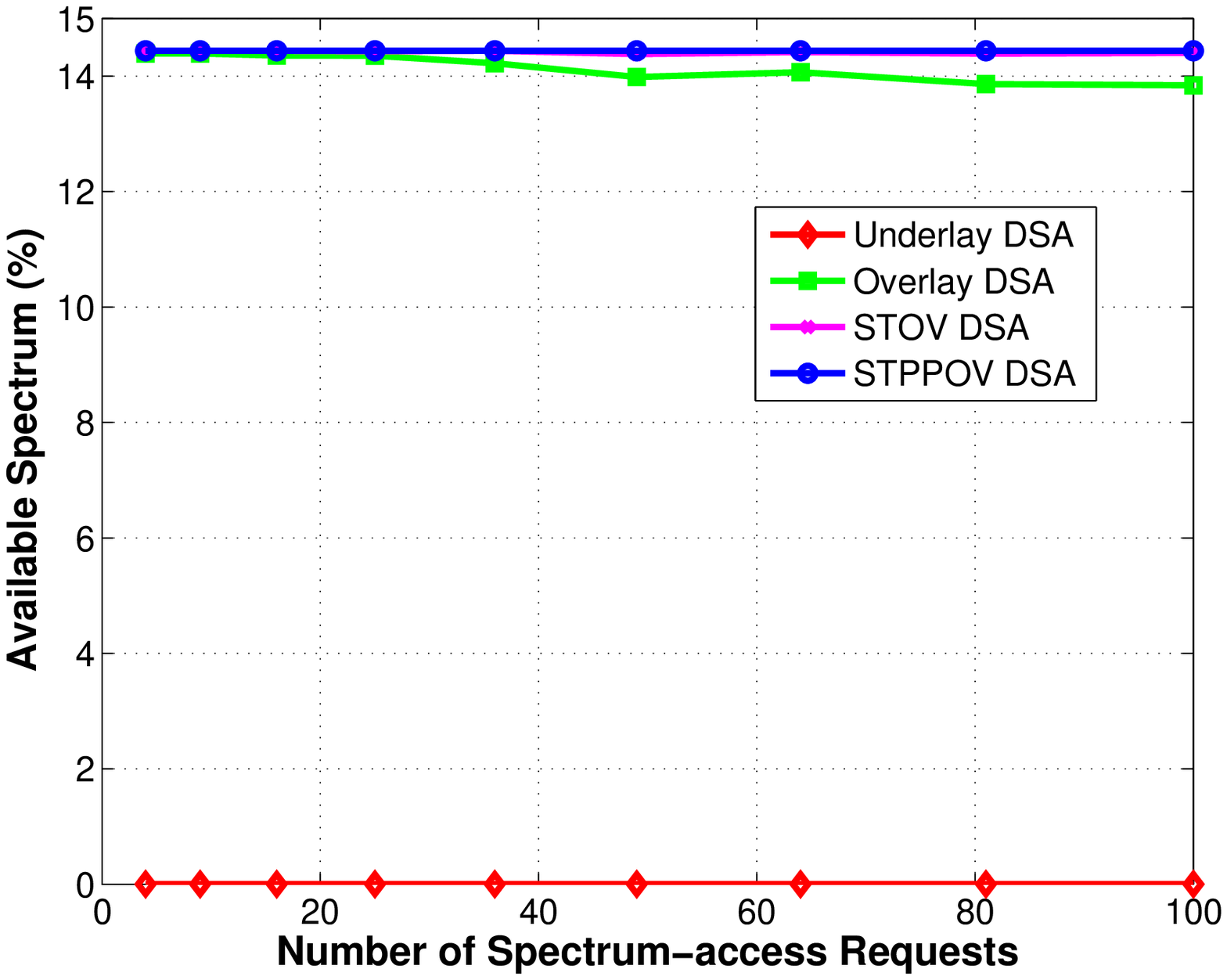}}
{\includegraphics [width=0.39\textwidth, angle=0] {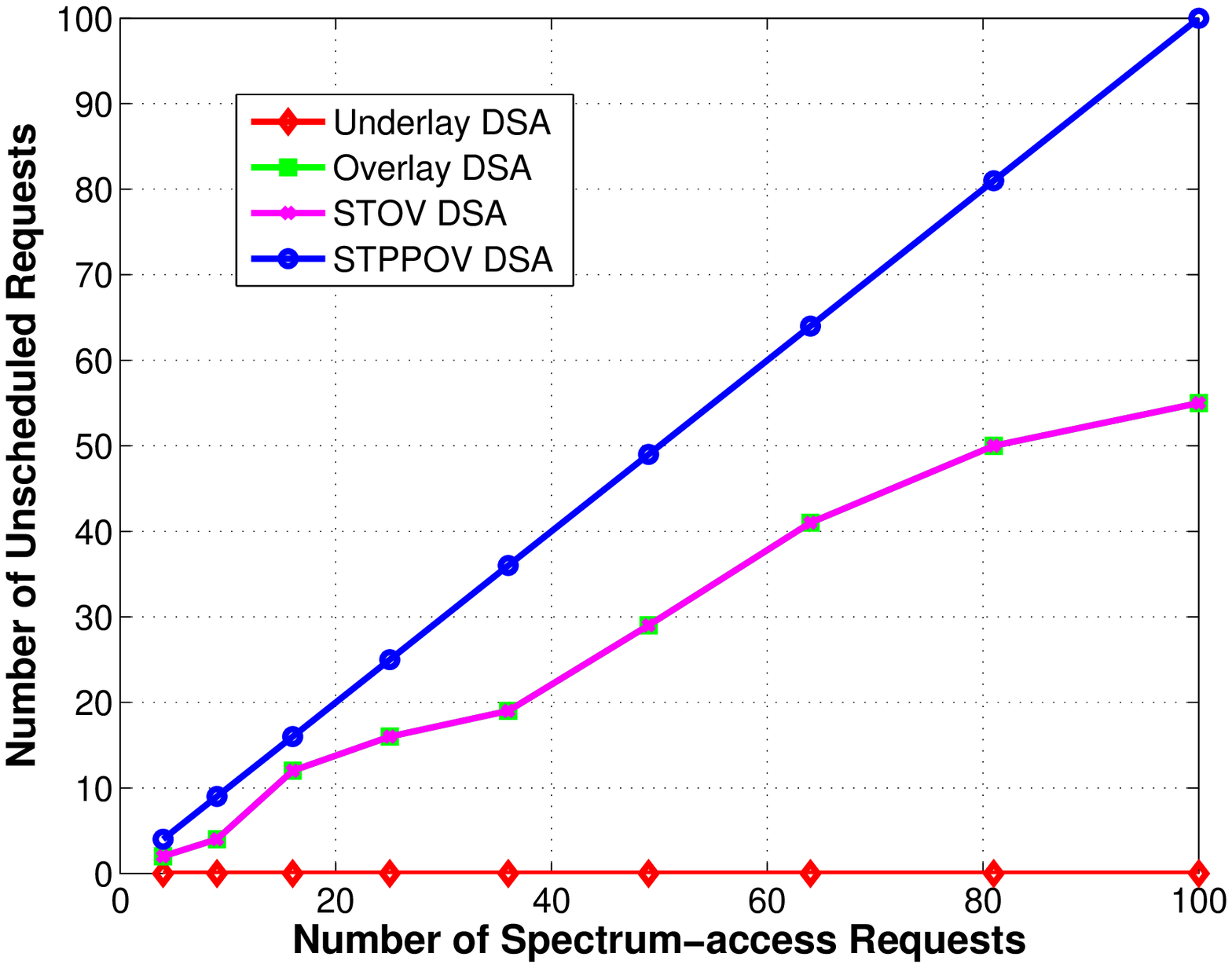}}
\caption{The performance comparison of four SAMs with varying number of the secondary networks when the PU is active. The PU network is operating. The worst-case PU receivers experience 10 dB higher SINR than the minimum desired SINR. The path-loss exponent is 3.5. The SU transceivers and PU receivers are assumed to be directional in order to minimize interference.}
\label{fig:LL501_ST5}
\end{figure}

The performance of the four SAMs for this base setup is shown in Figure \ref{fig:LL501_ST5}. We can see that:
\begin{itemize}
	\item In case of Underlay spectrum access, all the SU networks are exercising spectrum, thus the number of scheduled requests is same as the number of SU networks. However, since the transmit power is heavily constrained,  the signal power at the secondary receivers is very low. Also, these receivers experience interference from the PU transmitter and the transmitters of the cochannel SU networks.
	\item In case of Overlay spectrum access, when PU transmitter is not within the sensing range of the SU network transceivers, a spectrum access is exercised. The SU sensitivity is considered to be -80 dBm and the PLE of 3.5 which implies 1390 m of sensing range. Thus, very often, with Overlay spectrum-access policy, a spectrum-access is not performed and the spectrum consumption performance is poor when PU is present. When SU transmitter can exercise access to the spectrum, the signal power at the SU receivers may not be high due to constraints on the SU transmit-power and those receivers need to tolerate interference from the cochannel PU and SU transmissions.
	\item The dynamic Overlay (STOV) with unconstrained SU transmit power faces similar issues and shows low spectrum consumption performance when PU is present. Because it employs high transmit power, the SINR at the SU receivers is high and receiver spectrum consumption is low and the available spectrum remains high.
	\item The dynamic Overlay with protection of the PU receivers from aggregate interference (STPPOV) shows better performance as compared to overlay and dynamic overlay in the cases SUs can detect the presence of the PU transmitter but their transmissions do not cause harmful interference to the PU receivers. It, however, suffers with a lower spectrum sharing performance in general because of the guard margin to protect the PU receivers from the aggregate interference effect.
\end{itemize}

\section{Maximizing the Exploitation of Spectrum}
In this Appendix, we apply spectrum-space discretization for:
\begin{itemize}
	\item defining a scheduling strategy based on the spectrum consumption weights of the spectrum-access requests.
	\item choosing spectrum-access parameters or system design parameters in order to maximize spectrum exploitation.
\end{itemize}

The performance of spectrum sharing could be significantly improved if the incumbents play an active role in secondary spectrum access. We argue that significant amount of the available spectrum could be reclaimed if have the knowledge of the PU receiver positions. Secondly, in order to minimize the spectrum consumed by the primary receivers, the PU transmit-power needs to be increased.

Here, we perform an experiment where the incumbent playing an active role in secondary spectrum access. Here, we use the same setup from Appendix B but boost the transmit-power of the PU transmitter from 9 dB to 30 dB. We position the primary receivers at half the range that is 250 m. We avoid the assumption that the PU receivers at the worst-case positions and incorporate knowledge of the PU receiver positions. \textit{We add another candidate mechanism that performs scheduling based on the best-case spectrum consumption values (NSC-CX DSA)}. The problem of joint scheduling and power allocation is NP-hard \cite{jcpa_nphard}. NSC-CX is a suboptimal strategy to \textit{minimize the spectrum consumed}. It prioritizes the spectrum-access requests with lower \textit{minimal}\footnote{Minimal network spectrum consumption by a spectrum-access request occurs in the absense of cochannel interference.} spectrum consumption by its transceivers. First, we quantify the \textit{minimal} spectrum consumed by a candidate spectrum-access-request and next schedule spectrum-access-requests in the ascending order of \textit{minimal network spectrum consumption} while ensuring the minimum SINR constraint is satisfied for all the receivers. 
\begin{figure}[htbp!]
\centering
{\includegraphics [width=0.38\textwidth, angle=0] {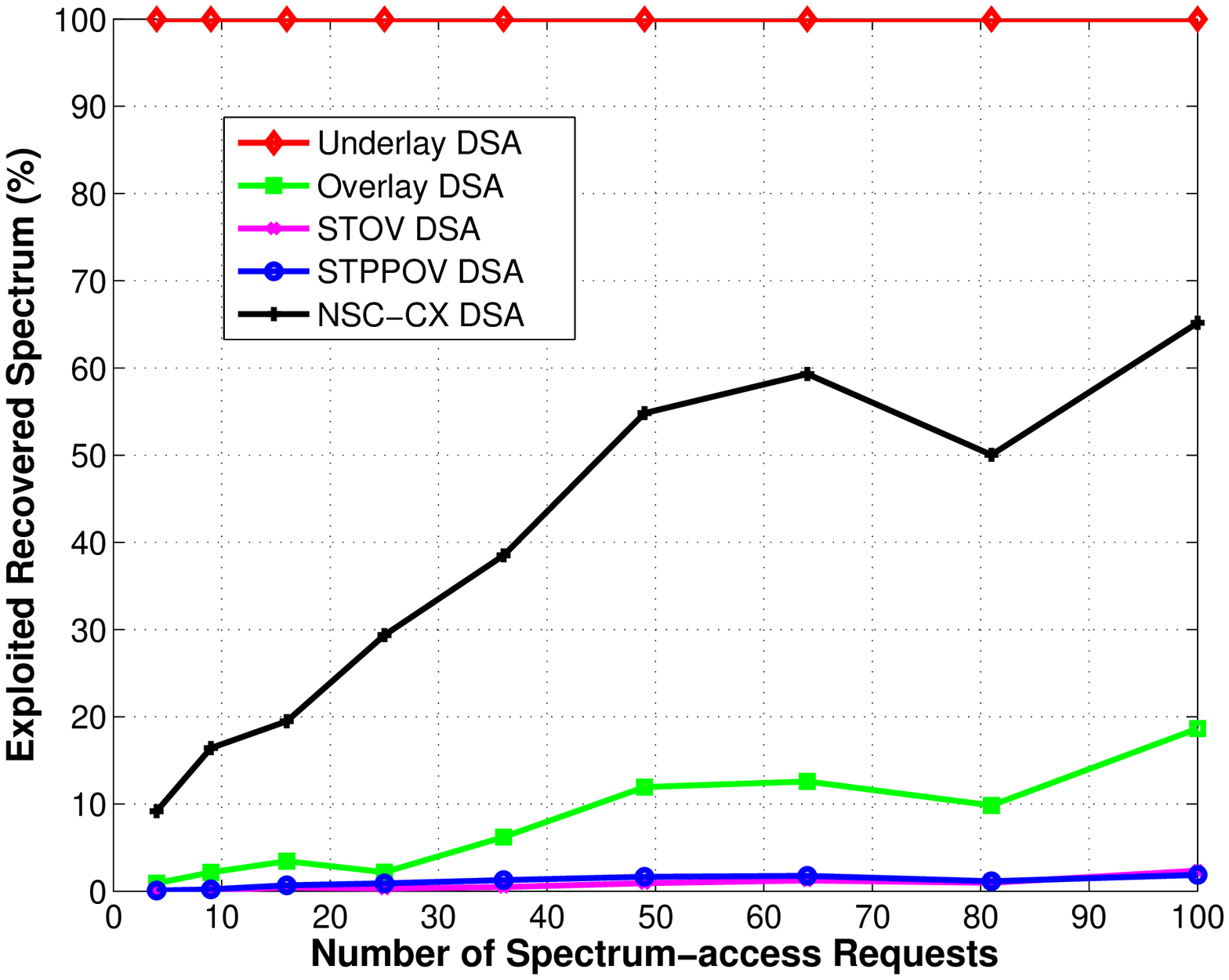}}
{\includegraphics [width=0.38\textwidth, angle=0] {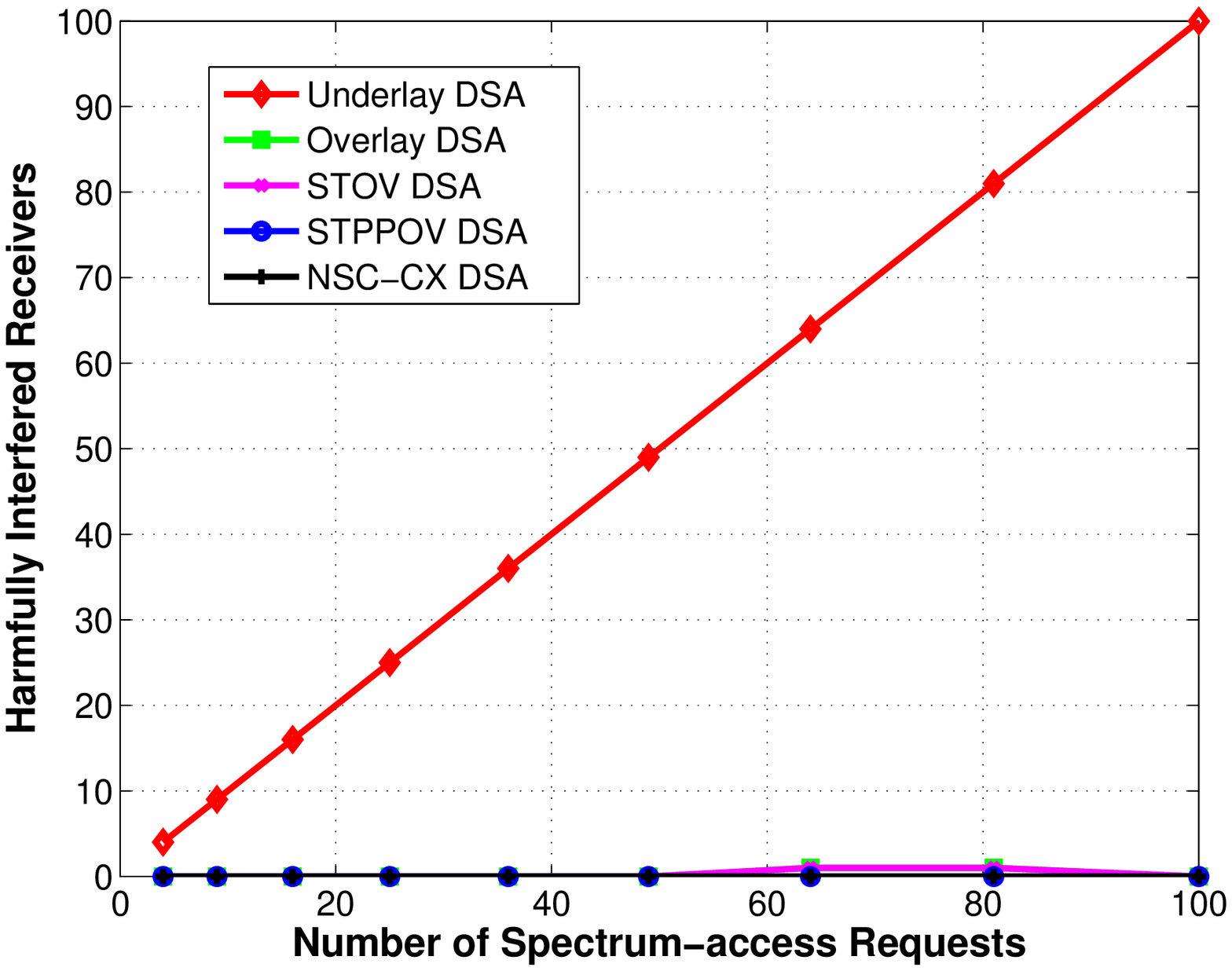}}
{\includegraphics [width=0.38\textwidth, angle=0] {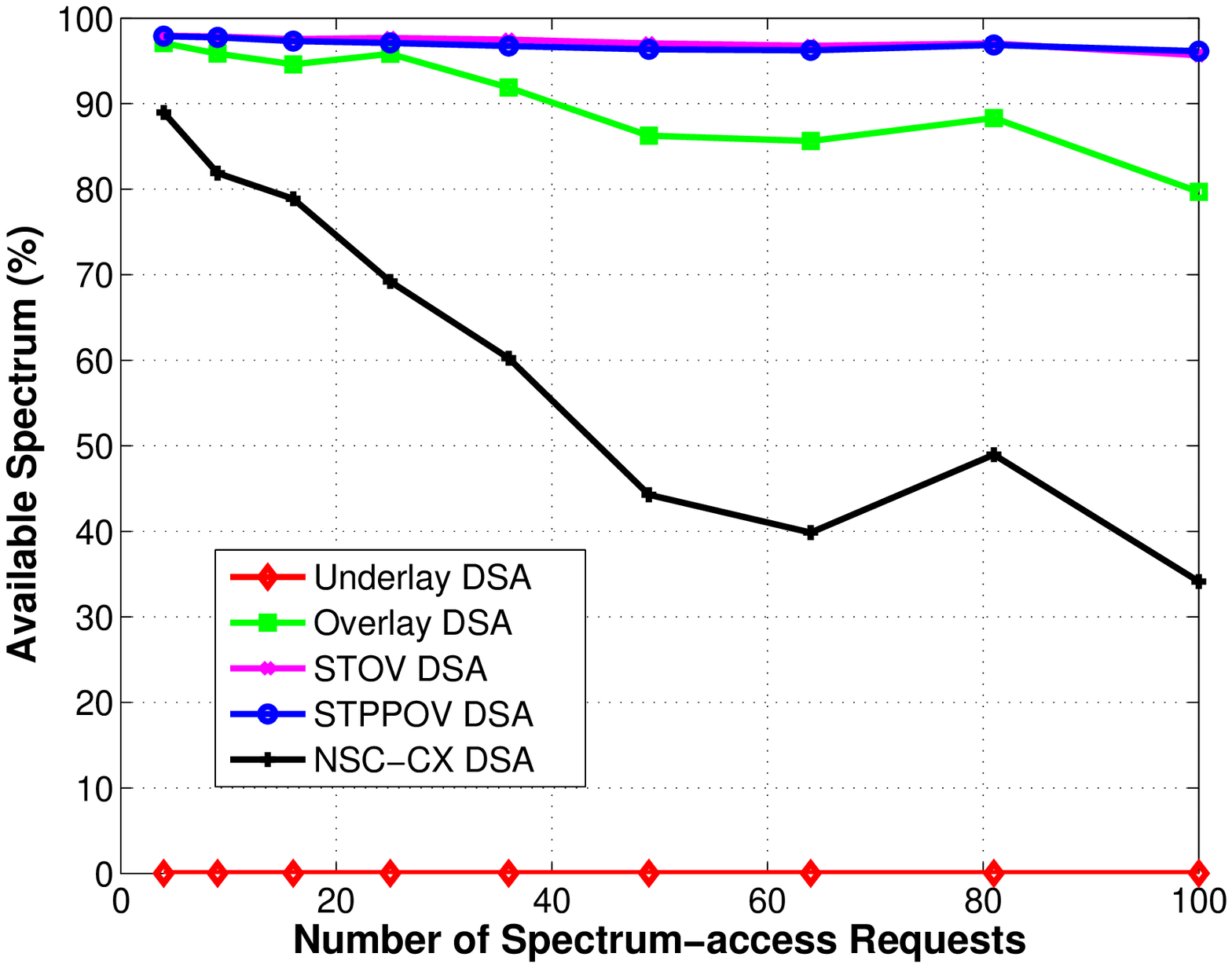}}
{\includegraphics [width=0.38\textwidth, angle=0] {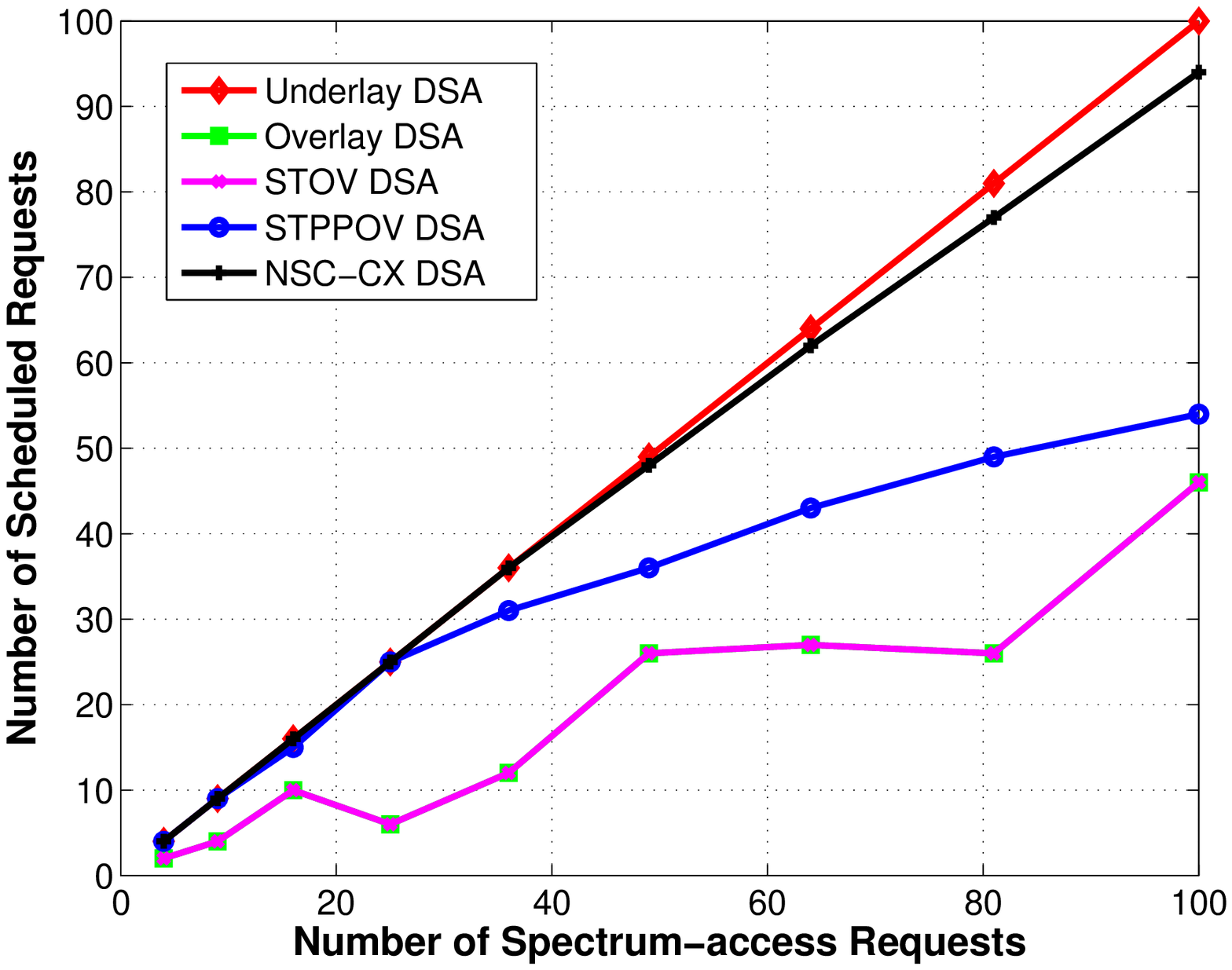}}
\caption{The performance comparison of SAMs with the varying number of the secondary networks when the incumbent is playing an active role. The PU transmit power is boosted in order to reduce the spectrum consumption by receivers. Also, the knowledge of the PU receiver positions is available to SUs in case of STPPOV-DSA and `NSC-CX DSA'. This results into higher amount of the available spectrum and significantly improves the performance of SAMs.}
\label{fig:LL501_ST7}
\end{figure}

\subsection{Effect of the Service Range of a Spectrum-access Request}
\begin{figure}[htbp!]
\centering
{\includegraphics [width=0.38\textwidth, angle=0] {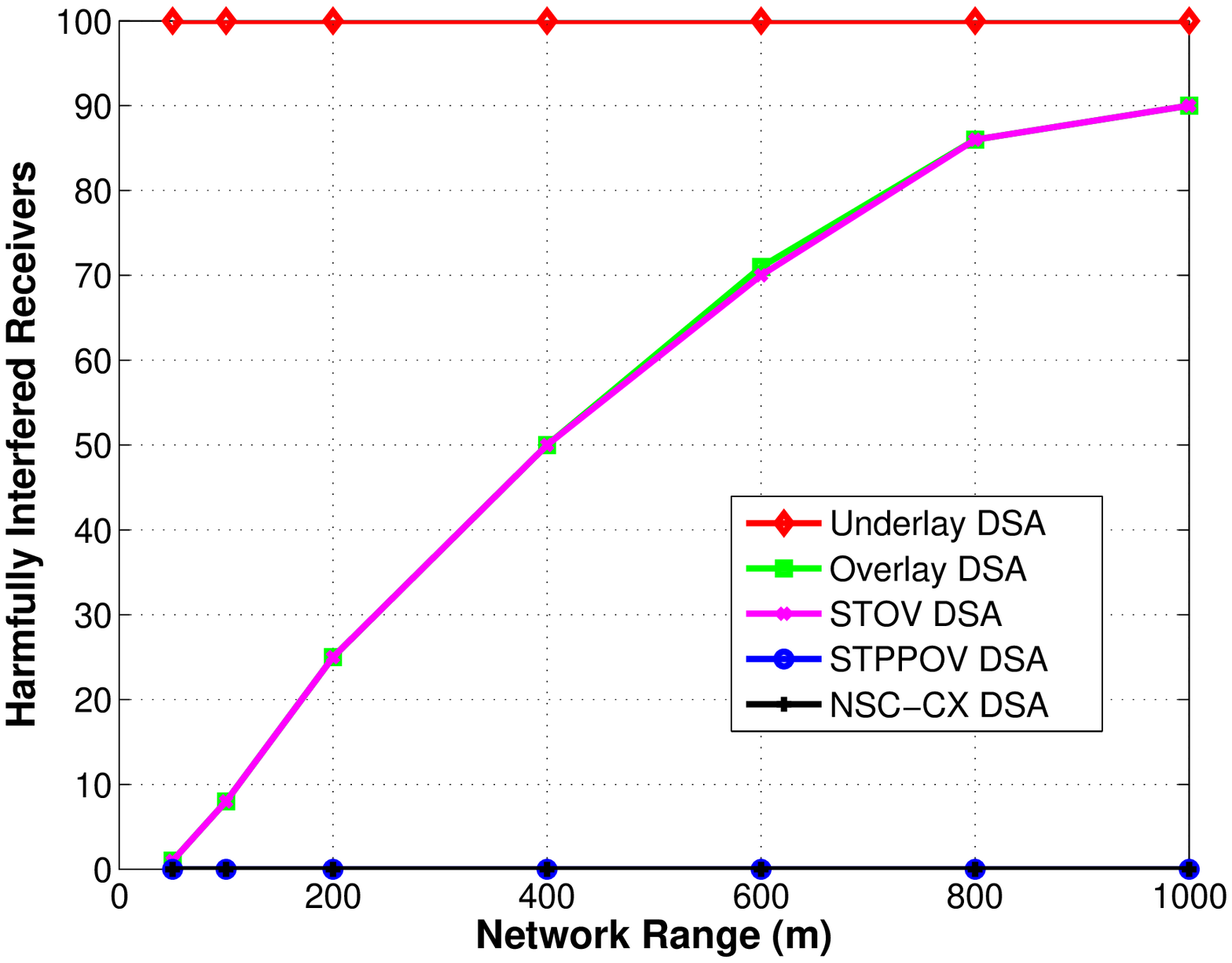}}
{\includegraphics [width=0.38\textwidth, angle=0] {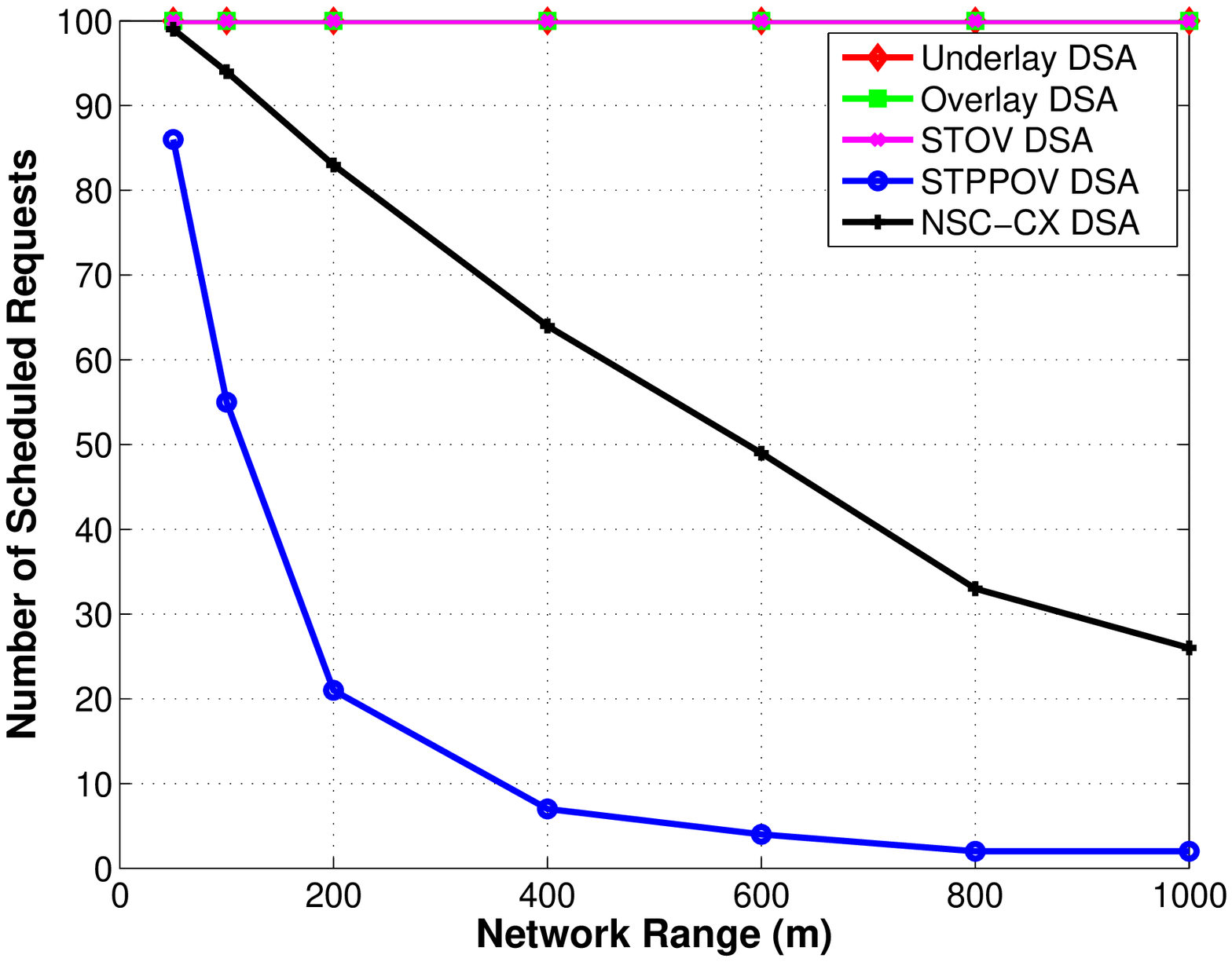}}
\caption{The performance comparison of SAMs with the varying network range. In this experiment, we observe that fine granular access with smaller network-ranges leads to higher spectrum-access requests being satisfied. }
\label{fig:L502_ST9}
\end{figure}

Next, we characterize the effect of service-range of the networks on the number of spectrum-access requests satisfied. From Figure~\ref{fig:L502_ST9}, we observe that The performance of `NSC-CX DSA' mechanism is reasonably good for all network ranges. Especially for the larger network ranges, `NSC-CX DSA' mechanism schedules a higher number of the spectrum-access requests as compared to other SAMs. This scenario helps us to understand the expected performance in terms of the number of successfully scheduled spectrum-access requests when larger network ranges are desired. With  close to 100 spectrum-access requests being satisfied for lower service-ranges, the experiment brings out the \textit{huge spectrum sharing potential with dynamic spectrum sharing}.

\section{Estimation of Spectrum Consumption Spaces}
To accomplish real-time dynamic spectrum sharing, it is important to estimate how much of the spectrum is used and how much is available for sharing. For estimating the used spectrum, we need to estimate the spatial distribution of \textit{spectrum-occupancy} in the geographical region as given by (\ref{eq:urspoc}). For estimating the available spectrum, we need to estimate the spatial distribution of \textit{spectrum-opportunity} as given by (\ref{eq:urspop}). 

Here, we resort to an external RF-sensor network that address the underlying sub-problems: signal detection, TDOA estimation, and received power estimation. The RF-sensor network is also used for characterize the propagation environment by estimating fine granular mean path-loss exponent and shadowing-loss. Exploiting signal cyclostationarity enables us to estimate the spectrum consumed by a transmitter in the presence multiple cochannel transmitters.

Based on the data acquired by RF-sensors, fusion center generates estimates for spectrum consumption estimates for each of the RF-sensors. In order to incorporate directional transmission, directional reception, and the propagation medium effects, it \textit{spatially} fuses each of the spectrum consumption spaces. 

\begin{figure}[htbp!]
\centering
{\includegraphics [width=0.31\textwidth, angle=0] {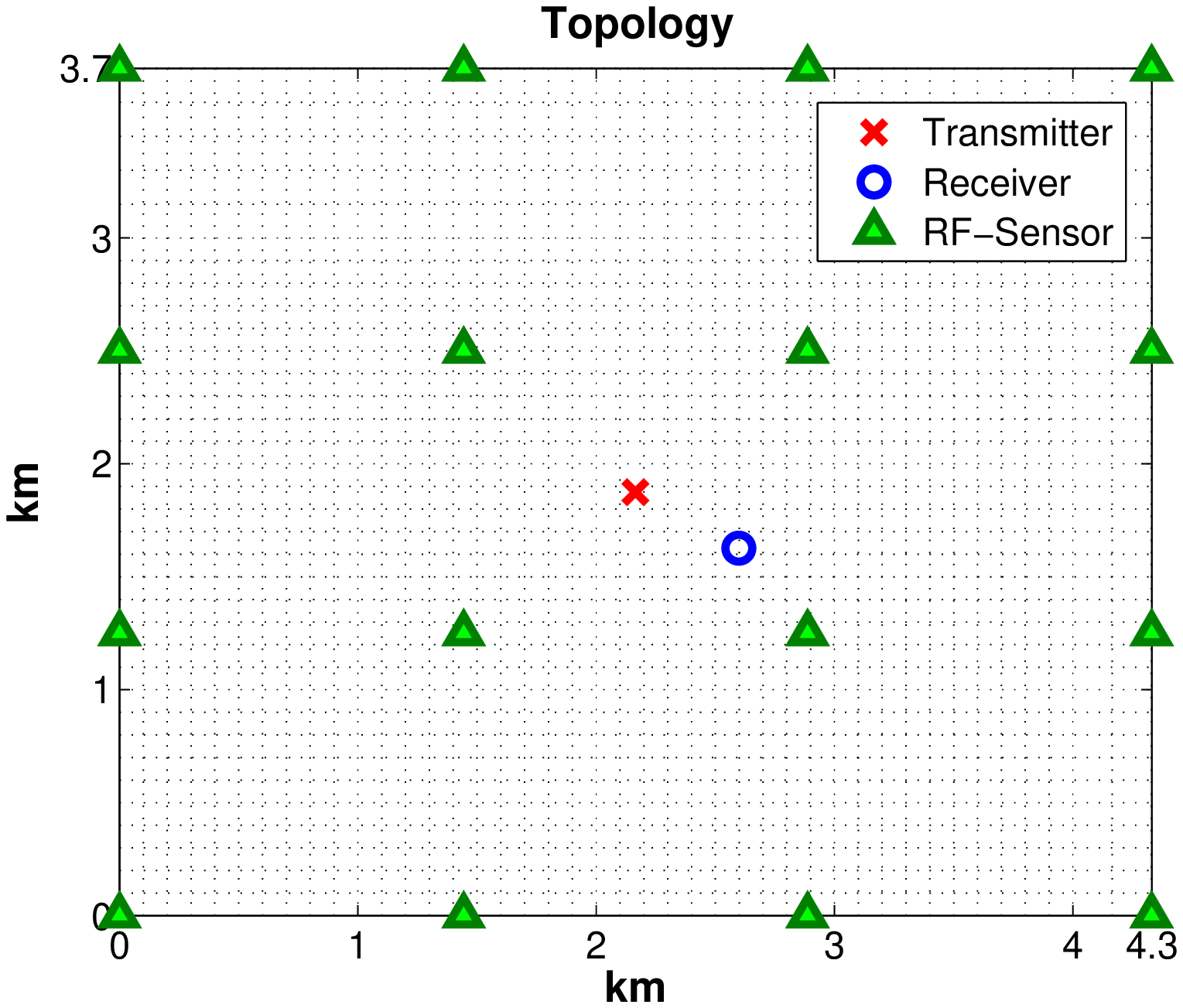}}
{\includegraphics [width=0.33\textwidth, angle=0] {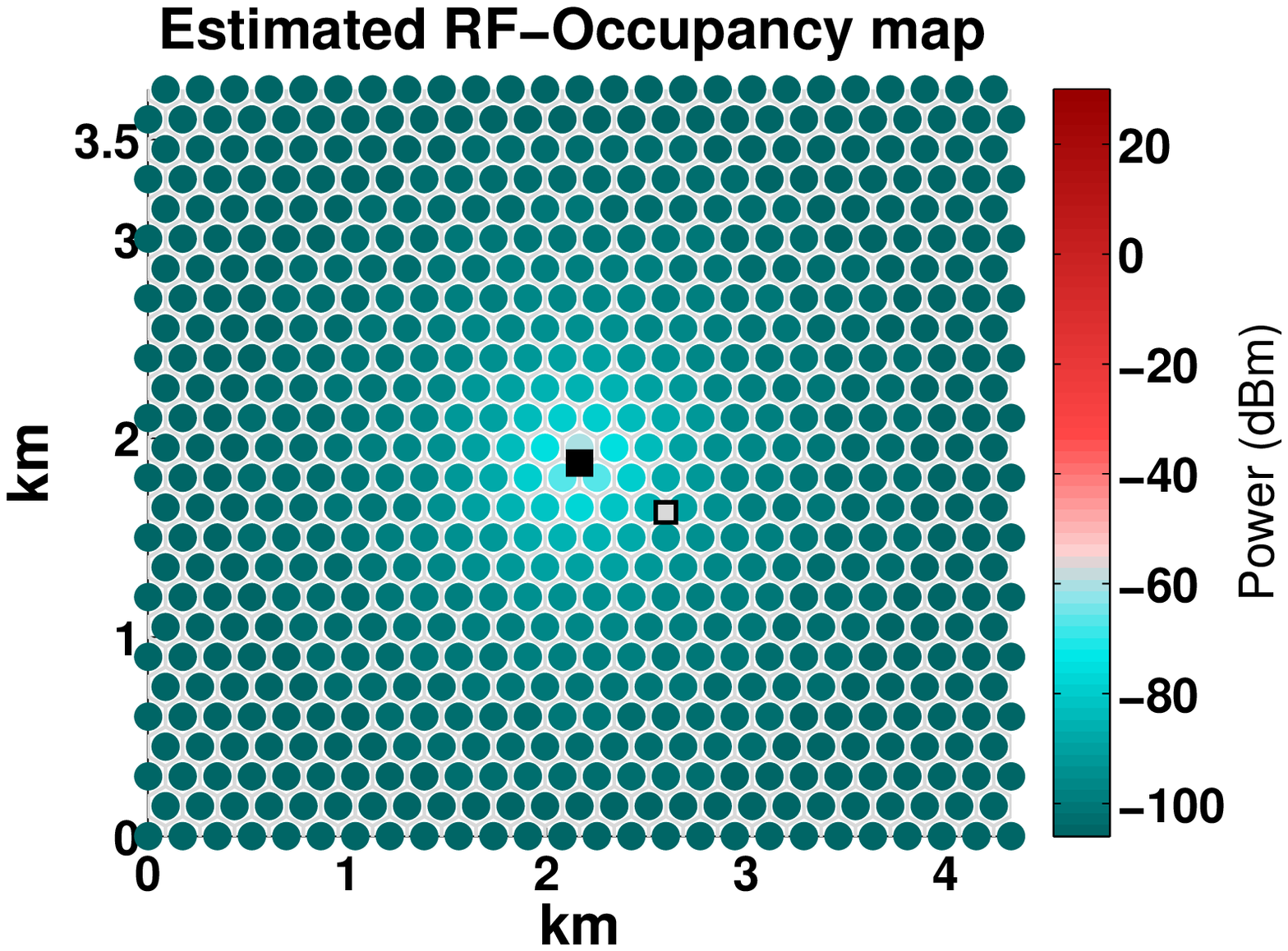}}
{\includegraphics [width=0.33\textwidth, angle=0] {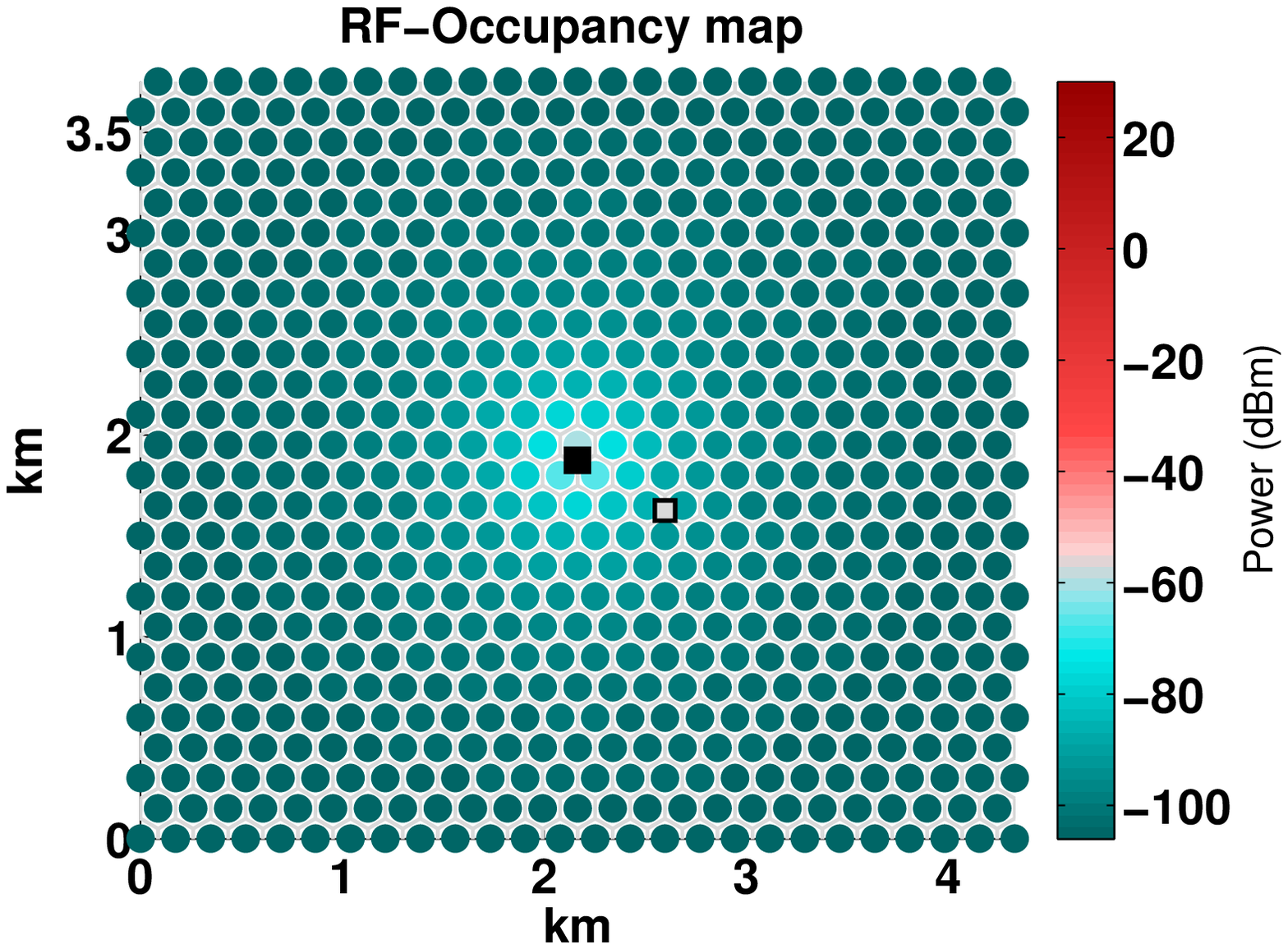}}
\caption{The figure shows the estimated and actual spectrum-occupancy maps for the topology shown on the left. The spectrum-occupancy map shows the spatial distribution of spectrum-occupancy in the units-regions of the geographical region that captures  the transmitter-utilized spectrum and }
\label{fig:B601_OCMAP}
\end{figure}

Figure \ref{fig:B601_OCMAP} shows a spectrum-occupancy map that captures the spatial distribution of the spectrum-occupancy in the unit-regions of the geographical region. 

\begin{figure}[htbp!]
\centering
{\includegraphics [width=0.33\textwidth, angle=0] {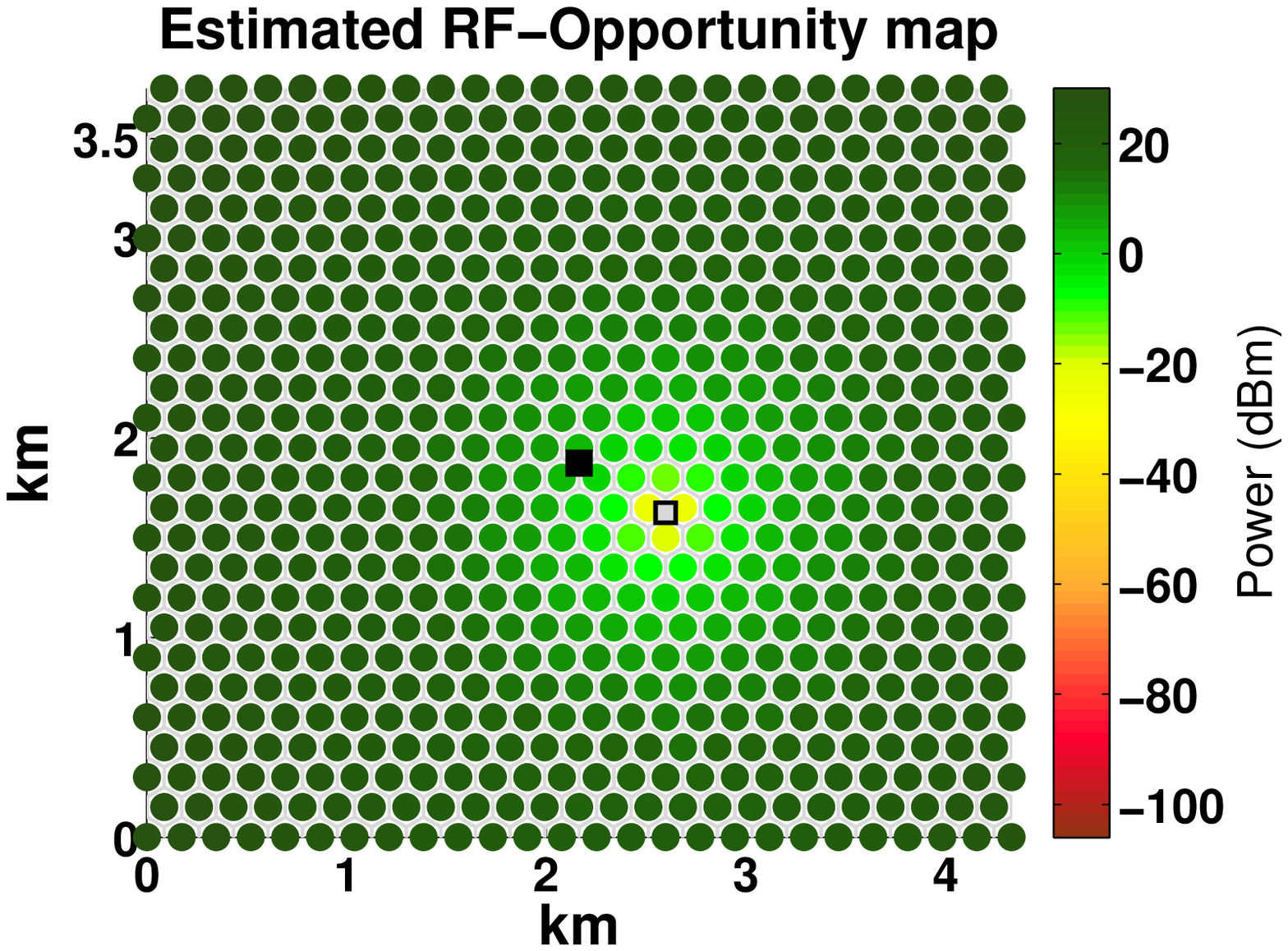}}
{\includegraphics [width=0.33\textwidth, angle=0] {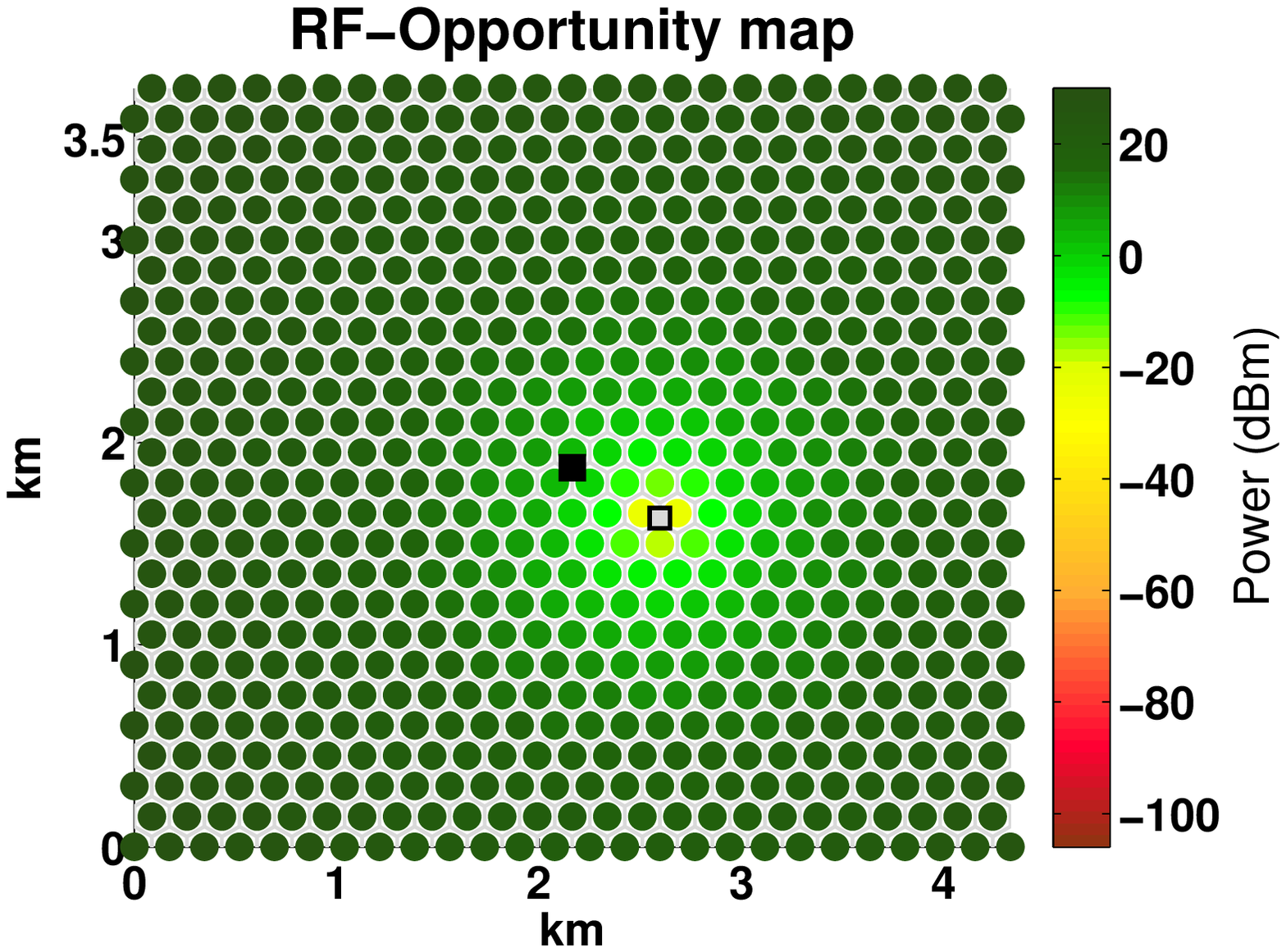}}
{\includegraphics [width=0.31\textwidth, angle=0] {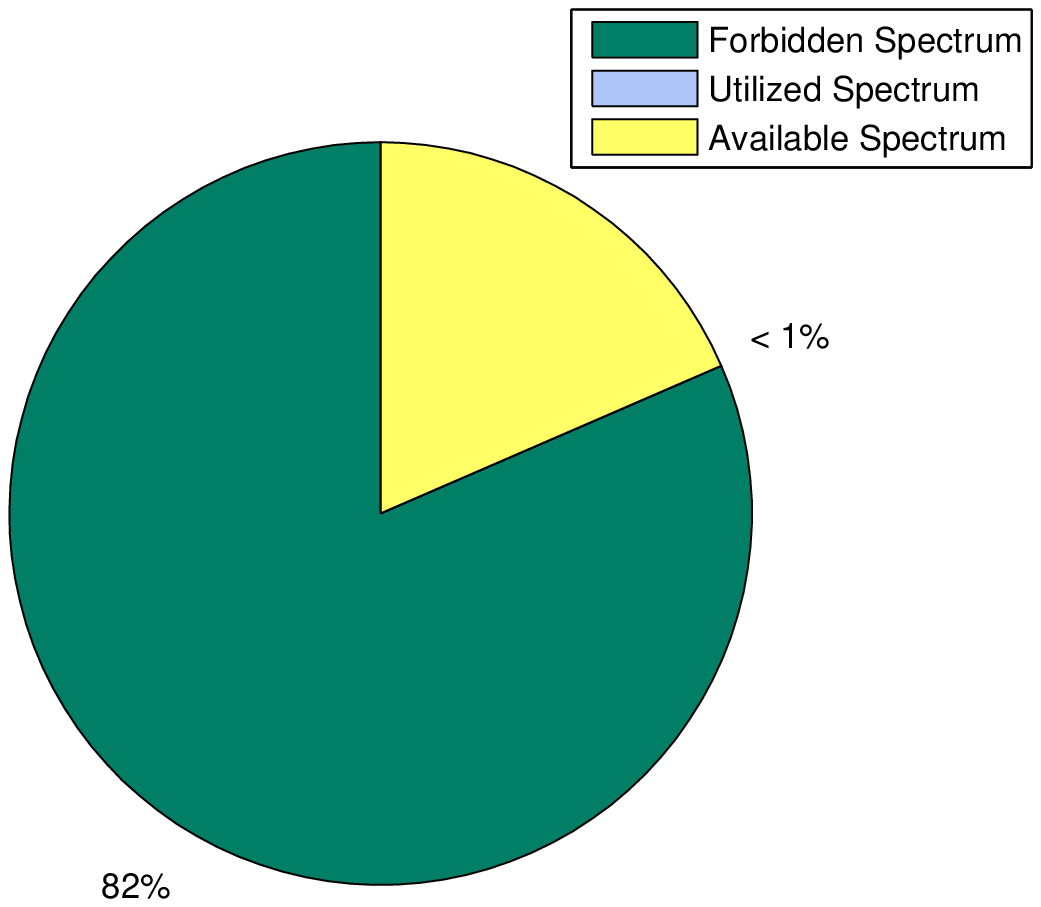}}
\caption{The figure shows the estimated and actual spectrum-opportunity maps. The spectrum-opportunity map shows the spatial distribution of  spectrum-opportunity in the units-regions of the geographical region and captures the available spectrum.}
\label{fig:B601_OPMAP}
\end{figure}

Estimating the SINR at the receivers enables us to estimate the interference-opportunity caused by a receiver.  The effective spectrum-opportunity in a unit-region is estimated using (\ref{eq:urspop}). Figure \ref{fig:B601_OPMAP} shows a spectrum-opportunity map that captures the spatial distribution of the spectrum-opportunity in the unit-regions of the geographical region.

To estimate the propagation environment, we use the simplest propagation model, the large scale path-loss model and apply it at a \textit{fine granular level}. We divide the geographical region into a grid of unit-sections. We deploy at least one RF-sensor within each unit section of the geographical region. We assume that a unit-section is small enough to have uniform shadowing. We model the propagation conditions in a unit-section in terms of \textit{mean path-loss exponent} and \textit{shadowing loss with respect to the mean path-loss exponent}. If the shadowing profile is to change at a much faster spatial rate, we need a large number of RF-sensors. When it is not possible to deploy very dense RF-sensor network, propagation environment could be characterized in the discretized spectrum space using satellite-maps or contour maps. Please refer to \cite{oms4_sce} for more details.

\end{document}